\documentclass[11pt]{article}
\usepackage[pdftex]{graphicx}
\usepackage{hyperref}
\usepackage{cite}
\usepackage{bm}
\usepackage{amsmath}
\usepackage{dsfont}
\usepackage{amsmath,amsfonts,amssymb,amsthm,nccmath,latexsym,mathtools}
\usepackage{xcolor}
\usepackage{amsmath}
\usepackage{cancel} %to strike out obliquely
\usepackage{amssymb}
\usepackage{amsbsy} % to write bold greek letters
\usepackage{footnote}

\hypersetup{
	colorlinks   = true, %Colours links instead of ugly boxes
	urlcolor     = blue, %Colour for external hyperlinks
	linkcolor    = blue, %Colour of internal links
	citecolor   = red %Colour of citations
}
\def\s{\sigma} 
\def\o{\omega} 
\def\O{\Omega}

\def\no{\nonumber}

\def\d{\delta}
\def\s{\sigma}

\def\p{\partial}

\def\na{\nabla}
\def\T{\Theta}
\def\th{\theta}
\def\k{\kappa}
\def\lie{\pounds_{\xi}}
\def\t{\tilde}
\def\kt{\t\kappa}
\def\thl{\theta_{(l)}}
\def\thlt{\t\theta_{(\t l)}}

\def\llt{\lambda_{(\t l)}}

\def\liek{\pounds_{\chi}}

\def\liel{\pounds_{l}}
\def\tliel{\pounds_{\t{l}}}

\def\be{\begin{equation}}
\def\ee{\end{equation}}
\def\ba{\begin{align}}
\def\ea{\end{align}}

\def\mg{\sqrt{-g}}
\def\tmg{\sqrt{-\tilde g}}

\def\ie{{\it i.e.~}}
\begin{document}
\title{{\bf{\Large Scalar-tensor gravity from thermodynamic and fluid-gravity perspective}}}
\author{
{\bf{\normalsize Krishnakanta Bhattacharya$^a$}}\footnote {krishnakanta@iucaa.in}\ \ \ and \ \
{\bf{\normalsize Bibhas Ranjan Majhi$^b$}}\footnote {bibhas.majhi@iitg.ac.in (Corresponding author)} \\
$^a$IUCAA, Post Bag 4, Pune University Campus,
   Ganeshkhind, Pune 411 007,
   Maharashtra, India.\\
$^b$Department of Physics, Indian Institute of Technology Guwahati, Guwahati 781039, Assam, India.}
\date{\today}
\maketitle
%%%%%%%%%%%%%%%%%%%%%%%%%%%%%%%%%%%%%%%%%%%%%%%%
\begin{abstract}
 Out of several possible extensions of general relativity, the scalar-tensor theory is the most popular for several reasons. Since the quantum description of gravity is yet to be formulated properly, the understanding of a gravitational theory remains incomplete until the study of thermodynamic and fluid-gravity aspects, which provides an alternative viewpoint to understand the gravitational theory. In this review, we study these features (thermodynamics and the fluid gravity analogy) in a rigorous and yet in a concise manner for the scalar-tensor gravity, which has been revealed in our recent works. In addition, the issue of conformally connected frames (\textit{i.e.} whether the two frames, which are conformally related are physically equivalent) has been explored in an explicit manner at the action level as well as from the viewpoint of thermodynamics and fluid-gravity correspondence.
\end{abstract}

%%%%%%%%%%%%%%%%%%%%%%%%%%%%%%%%%%%%%%%%

%%%%%%%%%%%%%%%%%%%%%%%

%\tableofcontents

%%%%%%%%%%%%%%%%%%%%%%%%%%%%%%%%%%%%%%%%%%%%%%

%%%%%%%%%%%%%%%%%%%%%%%%%%%%%%%%%%%%%%%%%%%%%%

%%%%%%%%%%%%%%%%%%%%%%%%%%%%%%%%%%%%%%%%%%%%%%%%%%%%%%%%%%%%%%%%%%%%%%%%%%%%%%%
\section{Introduction}
Among the four forces existing in the nature, the gravitational force is the oldest one to be discovered. However, the realization of the gravitation has been changed and has been modified with the course of time. The old Newtonian idea of gravitation was given in terms of ``inverse-squared'' law, where the gravitation was identified in terms of force acting upon two massive objects. This idea (of gravitation) lasted for several centuries until in 1916 when Einstein (along with due credit to Hilbert and others) came up with the idea of general relativity (GR), which paved a complete shift of paradigm to understand the gravity.  The present understanding of gravity (provided by GR) is premised upon the the idea of interplay of the spacetime geometry due to influence of the matter. Although, GR has been proven as highly accurate and consistent with the observational data \cite{Dyson:1920cwa, Will:2014kxa, Abbott:2016blz}, it is not quite compatible with another foundation of physics, \textit{i.e} the quantum mechanics. In other words, the quantum theory of gravity is not yet developed. Meanwhile, it has been found that the dynamical equations of GR show tantalizing similarity of the system of thermodynamics and fluid-dynamics. In the absence of a proper quantum theory of gravity, the thermodynamic and the fluid-dynamic aspects of gravity provide us with an alternative viewpoints to understand gravity.

In spite of huge success of GR, which we have discussed above, there are several motivations (both from theoretical and observational  viewpoints) to look beyond gravity. So far, the gravity has been tested in the weak field limit. It is argued that, in the strong gravity regime, the gravity can deviate substantially from GR and can be described by a modified theories of gravity. Among several modifications of GR, the scalar-tensor (ST) gravity is, probably, the most popular among the physicists for various reasons \cite{Callan:1985ia, EspositoFarese:2003ze, Elizalde:2004mq, Saridakis:2016ahq, Crisostomi:2016czh, Langlois:2017dyl}. This theory can be described in the two frames. In the original frame, which is also known as the Jordan frame, the gravity is described in therms of both the metric tensor and the scalar-field $\phi$, which is non-minimally coupled to the Ricci scalar $R$ in the action. The action of the Jordan frame can be written equivalently in a form (via the conformal transformation) which looks similar to the Einstein-Hilbert action along with the external scalar field. The latter action is known as the action of the Einstein frame. The physical equivalence of the Jordan and the Einstein frame has been a matter of debate over the years \cite{Faraoni:1999hp, ALAL,Faraoni:1998qx, Faraoni:2010yi, Faraoni:2006fx, Saltas:2010ga, Capozziello:2010sc, Padilla:2012ze, Koga:1998un, Jacobson:1993pf, Kang:1996rj, Deser:2006gt, Dehghani:2006xt, Sheykhi:2009vc,Steinwachs:2011zs, Kamenshchik:2014waa, Banerjee:2016lco, Pandey:2016unk, Ruf:2017xon,Karam:2017zno,Bahamonde:2017kbs,Karam:2018squ,Bhattacharya:2017pqc,Bhattacharya:2018xlq,Bhattacharya:2020wdl,Bhattacharya:2020jgk}.

 The main purpose of this review is providing the complete understanding of ST gravity from the alternative viewpoints \textit{i.e.} from the viewpoints of thermodynamics and fluid-gravity analogy. As we discuss later, the formulation of a consistent thermodynamic description in the two frames has been challenging and not been studied thoroughly earlier until our recent works \cite{Bhattacharya:2017pqc,Bhattacharya:2018xlq,Bhattacharya:2020jgk,Dey:2021rke}. In addition, recently we also have established the fluid-gravity analogy in ST gravity \cite{Bhattacharya:2020wdl}. Thus, we can now safely say that we now have a broader picture regarding the alternative ways of understanding scalar-tensor gravity than earlier. In this review, we tie up all the bits and pieces of information (which have been obtained in the course of time in several of our earlier works \cite{Bhattacharya:2017pqc,Bhattacharya:2018xlq,Bhattacharya:2020wdl,Bhattacharya:2020jgk,Dey:2021rke}) to provide the complete picture of thermodynamics and fluid-gravity analogy in ST gravity. In establishing these, it will be found that the validity of Padmanabhan's ``holographic relation'' (which was originally obtained for GR \cite{Padmanabhan:2002jr,Padmanabhan:2002xm,Padmanabhan:2003gd,Padmanabhan:2004fq,Padmanabhan:2006fn} and Lanczos-Lovelock gravity \cite{Mukhopadhyay:2006vu,Kolekar:2010dm,Kolekar:2011bb}) in ST gravity will provide a major clue to find the actual path. Moreover, providing a conformal transformation, a theory can be presented in infinite conformal frames. Therefore, the present analysis will help us to understand how (if at all!) the expression of thermodynamic and fluid parameters change under the conformal transformation. In addition, the understanding of thermodynamic and fluid-gravity aspects of ST gravity will also help us to properly understand those features of GR itself.
 
 The notations, which we follow here are the following: Any quantity with a tilde overhead (such as $\t A$) will correspond to the quantity of the Einstein frame. The quantities without a tilde will correspond to the same in the Jordan frame (such as $A$).

%%%%%%%%%%%%%%%%%%%%%%%%%%%%%%%%%%%%%%%%%%%%%%%%%%%%%%%%%%%%%%%%%%%%%%%%%%%%%%
\section{The scalar-tensor gravity: how, what \& why}
The attempt to describe gravity in terms of the scalar field can be traced back as early as 1913, when G. Nordstr$\ddot{\textrm{o}}$m came up with the idea of the scalar gravity \cite{NORDSTROM}. Later, this attempt was overshadowed by the seminal work of Einstein's general relativity (GR), where the interaction of gravity is carried by the second-ranked symmetric tensor field ($g_{ab}$, known as the metric tensor). A few years later, Kaluza \cite{KALUZA} and Klein \cite{KLIEN} proposed the classical unified field theory of gravitation and electromagnetism (known as KK theory), which was built in five dimension where the fifth dimension was considered to be a compact one. Among the fifteen independent components of the five-dimensional metric tensor, ten components were identified as the 4D metric tensor, four components were identified as the electromagnetic vector potential and the other one as the scalar field. With the assumption of the ``cylinder condition'' (\textit{i.e.} no component of the 5D metric depends upon the fifth dimension), it was shown that the 5D Einstein's equation yields the Einstein's equation in 4D along with the Maxwell's equation and the scalar field equation. Later, Jordan adopted the similar idea of KK theory and considered the four dimensional carved manifold to be embedded in a five-dimensional flat spacetime. He identified the fifteenth field variable of KK in terms of the gravitational constant, which fits nicely with the Dirac's idea of ``large number hypothesis'' \cite{Dirac:1938mt}. Later, in order to make Einstein-Hilbert gravity to be compatible with Mach's principal, Brans and Dicke took Jordan's idea of variable cosmological constant. Their proposed theory is known today as the Brans-Dicke theory \cite{Brans:1961sx}, which is a prototype of a more generalized scalar-tensor (ST) theory.

% five-dimensional theory of gravity with a compactified and constant fifth metric component. This Kaluza-Klein (KK) theory nicely unified the gravity and the electromagnetism. This is because, the KK theory provides the  Einstein's equation along with the Maxwell's equation in four dimensions.

 Unlike Einstein's gravity, the interplay of gravity in the spacetime in ST gravity is described not only in terms of the metric but also in terms of the scalar field, which is non-minimally coupled. The gravitational action of the ST theory in the original frame (a.k.a. the Jordan frame) is given as 

\begin{eqnarray}
&&\mathcal{A}=\int d^4x\sqrt{-g}L =\int d^4x\sqrt{-g} \frac{1}{16\pi}\Big(\phi R-\frac{\omega (\phi)}{\phi}g^{ab}\nabla_a\phi \nabla_b\phi -V(\phi)\Big)~.\ \
\label{SJ}
\end{eqnarray}
In the above action \eqref{SJ}, the scalar field $\phi$ is non-minimally coupled with the Ricci-scalar $R$. In addition, the coupling parameter $\o$, which is known as the Brans-Dicke (BD) parameter, is kept as an arbitrary function of the scalar field. When the BD parameter is constant, the ST theory boils down to the well-known Brans-Dicke theory. Furthermore, $V(\phi)$ is the arbitrary scalar field potential in the ST theory.
% Note, for simplicity we have not accounted the matter source or the matter Lagrangian. This is because, the presence of the external matter does not have much significance in the gravitational thermodynamics. Whenever the role of external matter alter the conclusions of the paper, we shall mention those cases properly.

 There is another representation of the above action \eqref{SJ}, which is known as the Einstein frame representation where the non-minimal coupling of the scalar field is no longer present. This is obtained by the following sets of transformation in the metric tensor and the scalar field. These are given as
\begin{align}
g_{ab}\rightarrow\tilde{g}_{ab}=\Omega^2g_{ab},\ \ \ \ \ \ \ \ \Omega=\sqrt{\phi} ~,
\label{GAB}
\end{align} 
and 
\begin{align}
 \phi\rightarrow\tilde{\phi}\,\ {\textrm{with}}\,\ d\tilde{\phi}=\sqrt{\frac{2\omega(\phi)+3}{16\pi}}\frac{d\phi}{\phi}~.
\label{PHI}
\end{align}
 The transformation relation, as described in Eq. \eqref{GAB}, is known as the conformal transformation, where $\Omega$ is the conformal factor. Due to this conformal transformation along with the rescaling of the scalar field $\phi$ (as mentioned in Eq. \eqref{PHI}), the Jordan frame action \eqref{SJ} can be written equivalently as

\begin{eqnarray}
&&\tilde{\mathcal{A}}=\int d^4x\sqrt{-\tilde{g}}\tilde{L}
=\int d^4x\sqrt{-\tilde{g}}[\frac{\tilde{R}}{16\pi}-\frac{1}{2}\tilde{g}^{ab}\tilde{\nabla}_a\tilde{\phi}\tilde{\nabla}_b\tilde{\phi}-U(\tilde{\phi})]~,
\label{SE}
\end{eqnarray}
with $U(\tilde{\phi}) = \frac{V(\phi)}{16\pi\phi^2}$. The above action \eqref{SE} corresponds to the Einstein frame representation of the ST gravity. Note, in the Einstein frame, the non-minimal coupling is no longer present and the scalar field appears as the external field.

%\subsection{Scalar-tensor gravity from observational viewpoints}
Before proceeding towards the main objective of this review let us mention briefly few observational aspects and recent progress of this theory.
As we have discussed earlier, ST theory is very popular from the observational viewpoints as well. Moreover, the ST gravity provides us with more free parameters, which can be fixed from the various observations; solar system tests, gravitational waves and from cosmology. Some important constraints in the parameters are discussed as follows.

Following the parametrized post-Newtonian (PPN) prescription (perturbative treatment of weak field gravity, for details about this formalism, see \cite{Quiros:2019ktw}), it is said that in the limit $\omega(\phi)\rightarrow\infty$, the ST theory boils down to the general relativity \cite{Nordtvedt:1970uv} and it has been a conviction for long that for strong coupling ($\omega$), the ST theory boils down to GR \cite{Billyard:1998kg,Barrow:1996kc,Mimoso:1995ge,Barrow:1994nx}. However, there are counterexamples of non-convergence in several cases \cite{Matsuda:1972zp,Romero:1992xx,Romero:1992bu,Romero:1992ci,Paiva:1993bv,Paiva:1993qa,Scheel:1994yn,Anchordoqui:1997du} and for the presence of scale-invariant matter fields (\textit{i.e.} for $T^i_i=0$) \cite{Banerjee:1996iy,Faraoni:1999yp}. Nevertheless, recent work agrees upon the convergence with GR even for $T^i_i=0$ at the quantum level \cite{Pal:2016hxt}. From the observation of the time delay of radio signal from the Cassini spacecraft, the constraint on the coupling parameter turns out as $\omega\geq 40000$ \cite{Bertotti:2003rm}.
It is also known that the scalar-tensor gravity predicts the velocity of gravitational wave as different from the velocity of light \cite{Bettoni:2016mij}. Interestingly, the recent observation of gravitational wave due to the binary neutron star merger (GW170817) is one with an electromagnetic counterpart (\textit{i.e.} GRB 170817A) and it was found that the electromagnetic signal was observed 1.7 s after the gravitational wave signal \cite{LIGOScientific:2017ync}. This event has provided stringent constraints on scalar-tensor gravity \cite{Baker:2017hug,Sakstein:2017xjx,Creminelli:2017sry}.
Thus, constraining the parameters in the scalar-tensor theory has been a subject of intense research over the years.

%{\color{blue}{%\subsection{Scalar hairs \& Spontaneous scalarization}
Apart from setting the constraints on the ST gravity (which arises from different avenues as discussed above), there are two other thriving directions which have recently surged attention among several researchers. These are, the study of scalar hairs and spontaneous scalarization, along with finding their observational implications. The usual ``no-hair theorem'' \cite{ruffini} of the black hole suggests that the black holes are characterized only by three physical quantities: mass, charge and the angular momentum. Thus, it does not matter what is the matter source, the final outcome of the gravitational collapse is always a Kerr-Newman black hole. However, while proving the no-hair theorem the focus is upon the dynamical end point of the gravitational collapse and not upon the existing black holes with arbitrary mass sources. Therefore, for existing black holes with non-trivial matter fields, later people found several other global charges or new non-trivial fields \cite{Bizon:1994dh,Bekenstein:1996pn,Volkov:1998cc}. These charges are categorized in two parts: (i) primary hairs, which are subject to the Gauss' law and are the global charges. These charges include mass, charge \textit{etc.}  (ii) secondary hairs, which are not the global charges and are not subject to the Gauss' law. However, irrespective of the mass source, the scalar hair, which are secondary hairs, was not found for the black holes. This lead Bekenstein to prove the ``no-scalar-hair theorems of black holes'' \cite{Bekenstein:1972ny,Bekenstein:1971hc,Bekenstein:1972ky}. However, while proving these theorem, Bekenstein have considered a few crucial assumptions \cite{Herdeiro:2015waa}. One such assumption was that the theorem is limited to those cases where the scalar field is canonical and is minimally coupled. However, this crucial assumption can be relaxed for various models of scalar-tensor gravity. Therefore, in that case, the scalar hairs are supposed to exist. During the last few years, the observational phenomena like the gravitational waves, observation of black hole shadow by Event Horizon Telescope (EHT) \textit{etc.} have provided new windows to test the gravity in the strong-field regime. Thus, checking validity of no-hair theorem from the observational phenomena has been one of the most important aspect to test gravity \cite{Isi:2019aib}. These has raised new interests on finding scalar hairs \cite{Hong:2020miv}, and thereby have raised interest in the theories where the scalar field is non-minimally coupled (such as the ST gravity \cite{Sotiriou:2013qea,Sotiriou:2014pfa,Khodadi:2020jij}).

Depending on the coupling, one can obtain solutions in ST gravity which are indistinguishable with GR in the weak gravity region but, they can predict radically different phenomenology in the strong gravity region, such as for the neutron stars and black holes.  One such phenomenology is referred to as the spontaneous scalarization. Originally this phenomena was found in the cosmological model introduced by Damour and Esposito-Far\`ese \cite{Damour:1993hw,Damour:1996ke}, where it was shown that a  tachyonic instability triggers the spontaneous scalarization and, at the end, the neutron star is dressed with a scalar configuration. This phenomenology has later explored in several cases \cite{Mendes:2016fby, SavasArapoglu:2019eil, Freire:2012mg,Esposito-Farese:2004azw,Ramazanoglu:2016kul,Yazadjiev:2016pcb,Liu:2020moh,Xu:2020vbs,Salopek:1988qh,Sennett:2017lcx,Barausse:2012da} and currently plays a crucial role from the observational viewpoints to obtain the deviation from Einstein's GR in the strong gravity region \cite{Zhang:2017sym,LIGOScientific:2018dkp,Shao:2017gwu,Anderson:2019eay,Guo:2021leu,Anderson:2016aoi}.

Let us now come back again to our theoretical study. Note that from the discussions presented in literature it seems that the two actions are exactly equivalent under the conformal transformation. However, a careful analysis shows \cite{Bhattacharya:2017pqc} that the two actions \eqref{SJ} and \eqref{SE} are equivalent only upto a total derivative term \textit{i.e.}

\begin{align}
\sqrt{-\tilde{g}}\tilde{L}=\sqrt{-g} L-\frac{3}{16\pi}\sqrt{-g}\square\phi~.
\label{ACT}
\end{align}
The origin of this total derivative term can be traced back from the following transformation relation of the Ricci scalar under \eqref{GAB}, which is provided as
\begin{align}
\tilde{R}=\frac{1}{\phi}\Big[R+\frac{3}{2\phi^2}(\nabla_i\phi)(\nabla^i\phi)-\frac{3}{\phi}\square\phi\Big]~. \label{RRCONF}
\end{align}
Since the $\square\phi$ term is a total derivative term, one can discard this term as removal of this term does not alter the dynamics of the system (follows from the fact that the equation of motion is unchanged irrespective of addition or removal of a total derivative term). This is why, for a long time $\square\phi$ has not been paid any attention and has been discarded. Therefore although there is an ``apparent physical equivalence'' at the dynamical level, the actions are indeed suffering from ``mathematical in-equivalence''. In the subsequent analysis, we shall show that this neglected term plays a crucial role in this theory and, therefore, it should be hailed high for its immense significance.

\subsection*{The conundrum of conformally connected frames}
 The issues with the conformal frames is, probably, the oldest one which has lingered with the theory ever since its formulation and has not been resolved yet. Providing the conformal transformation in the metric, a theory can be presented in infinitely different conformal frames among which the Jordan frame and the Einstein frame stand out. Now, there are the following issues related to the theory presented in two different conformal frame (say Jordan and Einstein frame).
 \begin{enumerate}
 \item Whether these different conformal frames, which are mathematically equivalent, are physically equivalent in every aspects.
 \item If the answer to the previous question (1) is negative then which frame is more physical than the others?
 \item If answer to the previous question (1) is positive then what should be the expression of the thermodynamic energy? In literature, there are several prescriptions of the definition of energy, which plays the role of thermodynamic energy in GR. There are Misner-Sharp energy, Hawking-Hayward quasi-local energy \textit{etc.} But, all these are not conformally invariant. Thus, if one agrees to the equivalent picture, one has to define the thermodynamic energy properly in a conformally invariant way and has to show the procedure to obtain the thermodynamic laws in a consistent manner.
 
 \end{enumerate}
 
 In the following analysis, we shall shed light on these issues. We shall show that the two frames are thermodynamically equivalent. In addition, at the action level, we shall show that in both the frames the surface terms and the bulk term are related by a special relation, known as the holographic relation. However, as we point out later on, the $\square\phi$ term plays the crucial role in both the cases.

%   It should be noted that 
%\begin{equation}
%\tilde{\mathcal{A}}_{GHY}=-\frac{1}{8\pi}\oint d^3x \sqrt{\tilde{h}}\tilde{K} \label{SEGHY}
%\end{equation}
%
%\begin{equation}
%\mathcal{A}_{GHY}=-\frac{1}{8\pi}\oint  d^3x \sqrt{h}\phi K~. 
%\label{SJGHY}
%\end{equation}
%
%\begin{equation}
%\tilde{K}=\frac{1}{\Omega}K-\frac{3}{\Omega^2}N^a\partial_a\Omega  \label{K}~,
%\end{equation}
%%%%%%%%%%%%%%%%%%%%%%%%%%%%%%%%%%%%%%%%%%%%%%%%%%%%%%%%%%%%%%%%%%%%%%%%%%%%%%
%\section{Decomposition of action as the bulk and the surface term: study of the holographic principle and equivalence/inequivalence at the classical level}

\section{Bulk and the surface terms: the holographic relation}

\subsection*{In search of a well-posed action principal}

The dynamics of a physical system is believed to be obtained via a well-posed action principle. However, in general relativity, there are several issues regarding the proper formulation of the action principle. Einstein's equation can be obtained from the Einstein-Hilbert (EH) Lagrangian, where the dynamical variable is usually considered as the metric tensor. Now, the EH Lagrangian contains the first order as well as the second order differentiation of the dynamical variable (\textit{i.e.} the metric tensor). Therefore, one has to fix both the metric as well as its first derivative on the boundary, which is inconsistent. To get rid of such trouble and to make the action principal well-posed, two routes are prescribed in the literature: (i) Addition of a boundary  term which, on the boundary, cancels the term containing the first derivative of the metric. The most popular boundary term in the literature is the Gibbons-Hawking-York (GHY) boundary term, which is dependent on the extrinsic curvature of the boundary surface under consideration. Therefore, this method is foliation dependent though generally covariant. (ii) The Einstein-Hilbert action is peculiar as the action as well as the derived field equation both contains up to the second derivative of the metric, which is against the general consensus (one expects the equation of motion to include the third derivative of the dynamic variable when the action itself contains its second derivative). This peculiarity is there because, all the second order derivatives in the Einstein-Hilbert action can be written as a total derivative as a whole. Thus, the action can be decomposed into two parts: the bulk part (which contains up to the first derivative of the metric tensor) and the surface part (which is a total derivative term and contains the second order derivatives). The equation of motion can be obtained only from the bulk part of the action. However, this method is not generally covariant. Furthermore, from the work of Padmanabhan \textit{et. al.} \cite{Padmanabhan:2003gd,Padmanabhan:2004fq,Mukhopadhyay:2006vu} a deeper significance of this decomposition was revealed as it was shown that the bulk part and the surface part are not independent. Instead, they are related via a relation which is call as the holographic relation (named by T. Padmanabhan). Therefore, if one of either bulk or the surface part is known, one can obtain another. In addition, it also suggests that Einstein's gravity is intrinsically holographic, where the surface degrees of freedom are related to the dynamics of the spacetime \cite{Bousso:2002ju}.

The above discussion has been presented under the framework of Einstein's gravity \cite{Padmanabhan:2002jr,Padmanabhan:2002xm,Padmanabhan:2003gd,Padmanabhan:2004fq,Padmanabhan:2006fn} and extended to Lanczos-Lovelock gravity \cite{Mukhopadhyay:2006vu,Kolekar:2010dm,Kolekar:2011bb}. However, the above two routes are required to be adopted for a well-defined action principle in ST gravity as well. Like the Einstein-Hilbert action, the actions of the ST gravity (in both the frames) contains the second derivative of the metric. In addition, the field equations also contain up to the second derivative of the fields. Thus the above arguments of the two routes are valid for ST gravity as well. In spite of the apparent similarity in the structure in the metric, the holographic relation, however, is not guaranteed especially due to the non-minimal coupling of $\phi$ in the action. In the following, we briefly discuss the two routes in the context of ST gravity. Finally, it will be shown that the holographic principle can be obtain for the ST gravity. However, for that, it will be shown that the $\square\phi$ term (of Eq. \eqref{ACT}) is required to be incorporated in the action of the Jordan frame.

The suitable GHY term for the actions (\eqref{SJ} and \eqref{SE}) are similar as of the Einstein's GR. The expression of the GHY terms in the two frames are provided as follows.
In the Jordan frame, the GHY term is given as
\begin{equation}
\mathcal{A}_{GHY}=-\frac{1}{8\pi}\oint  d^3x \sqrt{h}\phi K~. 
\label{SJGHY}
\end{equation}
Here $K=-\nabla_aN^a=\frac{1}{\sqrt{-g}}\partial_a(\sqrt{-g}N^a)$ is the trace of the extrinsic curvature tensor and $N^a$ is the unit normal to the boundary surface. Also, $h$ is the determinant of the induced metric corresponding to the boundary surface. Similarly, in the Einstein frame, the GHY term is provided as
\begin{equation}
\tilde{\mathcal{A}}_{GHY}=-\frac{1}{8\pi}\oint d^3x \sqrt{\tilde{h}}\tilde{K} \label{SEGHY}~.
\end{equation}
Interestingly, although the gravitational actions in the two frames (given by the equations \eqref{SJ} and \eqref{SE}) are equivalent only up to a total derivative term (see Eq. \eqref{ACT}), it can be shown that the sum of the gravitational action along with the GHY term, as a whole, are conformally invariant \cite{Bhattacharya:2017pqc}.
 
 We discuss the second route (for a well-posed action principal in ST gravity) in a more robust manner in the following as it caters crucial informations regarding the conformal equivalence of the two frames. First, we provide the decomposition of the Lagrangians (into bulk and surface terms) in the two frames. The detail result has been obtained in \cite{Bhattacharya:2017pqc}, and we quote those in the following. The Lagrangian in the Jordan frame, as described in Eq. \eqref{SJ}, can be written as decomposed into the bulk and surface part as $\sqrt{-g}L=\sqrt{-g}L_{bulk}+L_{sur}$ where
\begin{eqnarray}
&&L_{bulk}=(1/16\pi)\Big[\Omega^2g^{ab}[\Gamma^i_{ja}\Gamma^j_{ib}-\Gamma^{i}_{ab}\Gamma^j_{ij}]-2\Omega^2g^{ab}\Gamma^i_{ab}(\partial_i\ln\Omega)
+2\Omega^2\Gamma^i_{ij}(\partial^j\ln\Omega)\Big]
\nonumber
\\
&&-\frac{4}{16\pi}\omega\Omega^2(\partial_i\ln\Omega)(\partial^i\ln\Omega)-\frac{V(\phi)}{16\pi\phi^2}~,
\label{bulk}
\end{eqnarray}
and
\begin{eqnarray}
&& L_{sur}=\frac{1}{16\pi}\partial_c(\Omega^2\sqrt{-g}V^c)=(1/16\pi)\partial_c[\Omega^2\sqrt{-g}(g^{ik}\Gamma^c_{ik}-g^{ck}\Gamma^m_{km})]~,
\label{3}
\end{eqnarray}
where one can identify $V^c=g^{ik}\Gamma^c_{ik}-g^{ck}\Gamma^m_{km}$.
Similarly, the Lagrangian in the Einstein frame, as provided in the Eq. \eqref{SE}, can be decomposed into the bulk and the surface parts as

%\begin{align}
%\sqrt{-\tilde{g}}\tilde{R} = \sqrt{-\tilde{g}} \tilde{g}^{ab}(\tilde{\Gamma}^{i}_{ja}\tilde{\Gamma}^{j}_{ib}-\tilde{\Gamma}^{i}_{ab}\tilde{\Gamma}^{j}_{ij}) + \partial_{c}[\sqrt{-\tilde{g}}\tilde{V}^{c}]
%\label{DEC}
%\end{align}
%where, $\tilde{V}^{c}=\tilde{g}^{ik}\tilde{\Gamma}^{c}_{ik}-\tilde{g}^{ck}\tilde{\Gamma}^{m}_{km}$. Therefore, we can decompose the whole Lagrangian in the Einstein frame as $\sqrt{-\tilde{g}}\tilde{L}=\sqrt{-\tilde{g}}\tilde{L}_{bulk}+\tilde{L}_{sur}$, where we identify the bulk term in the Einstein frame is given by
\begin{align}
\tilde{L}_{bulk}=\frac{1}{16\pi}\tilde{g}^{ab}(\tilde{\Gamma}^{i}_{ja}\tilde{\Gamma}^{j}_{ib}-\tilde{\Gamma}^{i}_{ab}\tilde{\Gamma}^{j}_{ij}) -\frac{1}{2}\tilde{g}^{ab}\tilde{\nabla}_a\tilde{\phi}\tilde{\nabla}_b\tilde{\phi}-U(\tilde{\phi})~;
\label{QUAD}
\end{align}
and the surface term is given as
\begin{align}
\tilde{L}_{sur}=-\partial_c\tilde{P}^c~, 
\label{SUR}
\end{align}
with
\begin{align}
\tilde{P}^c=-\frac{1}{16\pi}\sqrt{-\tilde{g}}\tilde{V}^c=\frac{\sqrt{-\tilde{g}}}{16\pi}(\tilde{g}^{ck}\tilde{\Gamma}^i_{ki}-\tilde{g}^{ik}\tilde{\Gamma}^c_{ik})~, \label{PCTIL}
\end{align}
and $\tilde{V}^{c}=\tilde{g}^{ik}\tilde{\Gamma}^{c}_{ik}-\tilde{g}^{ck}\tilde{\Gamma}^{m}_{km}$.

The decomposition of the actions (of the two frames) in terms of the bulk and the surface part, which we have discussed above, has been done with the motivation to obtain a well-defined action principle. The surface terms of the decompositions in the two frame can be removed from the gravitational actions and one can safely obtain the equations of motion from the bulk part only (see our earlier work \cite{Bhattacharya:2017pqc} for details). One serious drawback in this approach (to obtain the dynamical equations from the bulk part only) is that this approach is not a covariant one as both the bulk and the surface part in the two frames are not covariant scalars. Nonetheless, this can be considered as a useful way to define a well-posed action principle. Then the question arises: Is there any importance of the surface part of the action? Apparently, it seems that the surface term of the decomposition has no significance and it just creates trouble in defining the well-posed action principal. However, the surface term is very significant and discloses the holographic nature of gravity. In the Einstein frame, it can be shown that the bulk and the surface part of the action are connected by the holographic relation, which is given as
\begin{align}
\tilde{L}_{sur}=-\partial_c\Big[\frac{\partial\sqrt{-\tilde{g}}\tilde{L}_{bulk}}{\partial\tilde{g}_{ij,c}}\tilde{g}_{ij}\Big]. \label{LEINTIL}
\end{align}
The above relation shows that the bulk and the surface part are not independent to each other. This relations also suggests that the surface degrees of freedom are related to the dynamical degrees of freedom, implying the holographic nature of gravity in the Einstein frame. The above relation \eqref{LEINTIL}, which implies the holographic nature of gravity, also suggests that the gravitational action in the Einstein frame (given by Eq. \eqref{SE}) can be described as the action in the momentum space for the following reasons. Let us consider a pair of Lagrangians of the following forms: $L_1\equiv L_1$($q^A$, $\partial q^A$) and $L_2\equiv L_2$($q^A$, $\partial q^A$, $\partial^2q^A$). If $L_1$($q^A$, $\partial q^A$) and $L_2$($q^A$, $\partial q^A$, $\partial^2q^A$) are connected to each other by the relation $L_2$($q^A$, $\partial q^A$, $\partial^2q^A$)=$L_1$($q^A$, $\partial q^A)-\partial_i(q^Ap^i_A)$, it can be shown that both $L_1$($q^A$, $\partial q^A$) and $L_2$($q^A$, $\partial q^A$, $\partial^2q^A$) yields the same equation of motion; where in the first case (\textit{i.e.} for $L_1$($q^A$, $\partial q^A$)) one has to fix the coordinates $q^A$ on the boundary and in the second case, one has to fix the conjugate momenta ($p^i_A=\p L_1/\p(\p_iq^A)$) on the boundary. Thus, under the light of the above discussions, $\tilde{L}_{bulk}$ can be interpreted as the Lagrangian of the coordinate space and, on the other hand the total Lagrangian in the Einstein frame can be interpreted as the Lagrangian of the momentum space.

In spite of the fact, that the actions in the two frames are equivalent to each other, it can be shown that the holographic principle does not hold in the Jordan frame and the total Lagrangian in the Jordan frame cannot be interpreted as one of the momentum space. In the Jordan frame, the bulk part and the surface part of the Lagrangian are related to each other by the following relation:
\begin{eqnarray}
&&L_{sur}=-\p_c\Big[\frac{\partial\sqrt{-g}L_{bulk}}{\partial g_{ab,c}}g_{ab}\Big]+\frac{3}{16\pi}\sqrt{-g}\square\phi~. \label{EQU}
\end{eqnarray}
Therefore, it can be said that the holographic relation does not hold for the Jordan frame, described by the Lagrangian \eqref{SJ}. Furthermore, in this case, the Lagrangian $L$ cannot be interpreted as the action of the momentum space.
Since the holographic relation breaks down in the Jordan frame of ST gravity, there lies an in-equivalence even at the classical level. Later, we shall see, this in-equivalence will transcend at the thermodynamic level as well, making the thermodynamic parameters not to be exactly equivalent in the two frames. However, the reason for this in-equivalence can be traced back to the earlier relation \eqref{ACT}, where it has been shown that the actions in the two frames are equivalent only upto a total derivative term. To ward off this in-built in-equivalence in the two frames, we incorporate the $\square\phi$ term in the Lagrangian and define the Lagrangian in the Jordan frame as
\begin{align}
 L'=L-\frac{3}{16\pi}\square\phi~. \label{L'}
\end{align}
Now, it is quite straightforward to show that the actions in the two frames are exactly equivalent i.e., $\mathcal{A}'=\int d^4x\sqrt{-g}L'=\int d^4x\sqrt{-\t g}\t L=\tilde{\mathcal{A}}$~. Let us now check whether we can now establish the holographic relation. Note that the Lagrangian $L'$ can also be decomposed in terms of the bulk part and the surface part i.e. $\sqrt{-g}L'=\sqrt{-g}L'_{bulk}+L'_{sur}$, where the bulk part $L'_{bulk}=L_{bulk}$, is given by the relation \eqref{bulk} and the surface term will be given as 
\begin{align}
L'_{sur}=L_{sur}-\frac{3\mg}{16\pi}\square\phi=\frac{1}{16\pi}\partial_c\Big[\sqrt{-g}\Big\{\phi(g^{ik}\Gamma^c_{ik}-g^{ck}\Gamma^m_{km})-3g^{cd}\p_d\phi\Big\}\Big]~.
\end{align}
One can check that the holographic relation in this frame can be established as the bulk part and the surface part of the Lagrangian are related to each other as 
\begin{align}
L'_{sur}=-\p_c\Big(\frac{\partial\sqrt{-g}L'_{bulk}}{\partial g_{ab,c}}g_{ab}\Big)~. \label{HOLNEW}
\end{align}
Thus, with the modification of the Lagrangian in the Jordan frame, we have now removed the inequivalence of the two frames and have established the holographic relation in the Jordan frame. In addition, the above relation \eqref{HOLNEW} helps us to identify the bulk part of the Lagrangian (\textit{i.e.} $L'_{bulk}$ or $L_{bulk}$) as the Lagrangian of the coordinate space and the Lagrangian $L'$ as the same of the momentum space.

  Before proceeding further, we clarify again that the surface terms discussed in Eq (\ref{SUR}) and in Eq (\ref{PCTIL}) are not the same as the Gibbons-Hawking-York (GHY) term. Also see section 6.2.3 of the ref \cite{gravitation} where the difference of these two has been explicitly shown for a generic metric. In order to obtain a well-posed action principle, two roots are adopted. In one route, a total derivative term, such as the GHY terms is added along with the gravitational action (such as the EH action). In the second route, the problematic part of the gravitational action, \textit{i.e.} the surface pert (defined in Eq (\ref{SUR}) and in Eq (\ref{PCTIL}), which contains second order derivative of the metric) is discarded from the gravitational action and the equation of motion is obtained from the bulk part only (for details, see \cite{gravitation}). Also, note that there are numerous surface terms which can be added with the EH action instead of the GHY term, which most popular among those several alternatives (see ref \cite{Charap:1982kn} for a detail discussion). Furthermore, the GHY term crucially depends on the foliation of the spacetime and acts as a surface term only for the timelike or for the spacelike boundaries and does not work for a null surface (for the prescription on the null surface, see \cite{Parattu:2015gga,Parattu:2016trq,Chakraborty:2016yna,Jubb:2016qzt,Chakraborty:2018dvi}). On the other hand, the surface term of the action (discussed in Eq (\ref{SUR}) and Eq (\ref{PCTIL})) works in every of those cases, though the second method is not a covariant one. Thus, the surface part of the EH action and the GHY term are significantly distinct from one another.
  
  Now we clarify another terminology, which we have referred to as the ``holographic principle/relation''. This terminology of ``holographic principle/relation'' implies different meaning compared to the ``holographic principle'' arising from the AdS-CFT/ string theory.  The terminology which we use, was coined by Padmanabhan when he had shown that the bulk part and the surface part of the EH action (as described in the context of the second method) are not independent. Instead, these two are intrinsically connected to each other, which he termed as the ``holographic principle/ relation'' (see ref \cite{Padmanabhan:2004fq}). We know that the bulk part of the EH action contributes to the dynamics of the gravitational system as one can obtain the Einstein's equation from the bulk part alone (following the second method which we have discussed above). Now, since the bulk part of the action is connected to the surface part via a relation, the surface degrees of freedom contributes to the dynamics of the gravitational system. Moreover the bulk information can be achieved from the surface. In this sense the obtained relation was initially called as holographic relation  (see \cite{Padmanabhan:2004fq} for his comments on this ``holographic principle''). We also have used the same nomenclature here as well. In this regard it may be noted that Padmanabhan's (as well as ours) discussion on holographic principle is not the same as the concept of holographic principle provided by the string theory or AdS/CFT. 

In the following section, it will be shown that the thermodynamic parameters, obtained from the Lagrangian $L'$ are exactly equivalent to the same of the Einstein frame. Thus, from now on, we shall consider the Lagrangian of the Jordan frame as $L'$ instead of $L$.

%%%%%%%%%%%%%%%%%%%%%%%%%%%%%%%%%%%%%%%%%%%%%%%%%%%%%%%%%%%%%%%%%%%%%%%%%%%%%%%%%%
\section{Conserved Noether and ADT currents}
In thermodynamic structure of gravitational theories, the conserved currents play a crucial role in defining the thermodynamic parameters. In addition, the conserved currents like the Noether current due to diffeomorphism symmetry of gravity and the Abbott-Deser-Tekin (ADT) current provide the thermodynamic laws in a covariant way. In the following, it will be shown that the Wald's formalism \cite{Wald:1993nt} and the ADT formalism \cite{Abbott:1981ff,Deser:2002rt, Deser:2002jk, Deser:2003vh}, which are based upon the Noether current and the ADT current respectively, will be useful in order to obtain the 1st law in a covariant way. Also, it will be shown that the thermodynamic parameters are equivalent in the two frames. Since we already have the ambiguity regarding the potential candidate for the internal energy in thermodynamics of ST gravity, this Wald's formalism (or equivalently the ADT formalism) will be proved to be highly useful in defining the thermodynamic energy, which is also conformally invariant and fits nicely with other thermodynamic parameters to provide the first law. In the following, we show the procedure to obtain the Noether and the ADT currents in the two frames. To obtain the Noether current, we firstly obtain the variation of the Lagrangians in the two frames. Thereby we obtain the equations of motion and the boundary terms. Later we specify this arbitrary variation as the change due to the diffeomorphism. Therefore, the arbitrary variations in the Lagrangians will be changed to the Lie variation. This will yield the expression of the Noether current due to the diffeomorphism invariance.
\subsection{Variation of the actions in the two frames, equations of motion and the boundary terms} \label{var}

The arbitrary variation of the Lagrangian in the Jordan frame yields
\begin{align}
\delta(\sqrt{-g}L')=\sqrt{-g}\Big(E_{ab}\delta g^{ab}+E_{(\phi)}\d\phi+\na_a\T'^a(q,\d q)\Big)~, \label{DELL'}
\end{align}
where $q\in \{g_{ab}, \phi\}$ and 
\begin{eqnarray}
&& E_{ab}=\frac{1}{16\pi}[\phi G_{ab}+\frac{\omega}{2\phi}\nabla_i\phi\nabla^i\phi g_{ab}-\frac{\omega}{\phi}\nabla_a\phi\nabla_b\phi
+\frac{V}{2}g_{ab}-\nabla_a\nabla_b\phi+\nabla_i\nabla^i\phi g_{ab}]~;
\no
\\
&& E_{(\phi)}=\frac{1}{16\pi}[R+\frac{1}{\phi}\frac{d\omega}{d\phi}\nabla_i\phi\nabla^i\phi +\frac{2\o}{\phi}\square\phi-\frac{dV}{d\phi}-\frac{\omega}{\phi^2}\nabla_a\phi \nabla^a\phi]~;
\no 
\\
&&\ \ \ \ \ \ \ \ \ \ \ \ \ \ \ \ \ \ \ \ \ \ \ \ \ \ \ \ \ \ \ \ \ \ \ \ \ \ \ \textrm{and} 
\no 
\\
&&\T'^a (q,\delta q)=\T^a (q,\delta q)-\frac{1}{16\pi}\Big\{\frac{3}{2}g^{ij}\d g_{ij}\p^a\phi-3g^{ia}\p^b\phi\d g_{ib}+3\p^a(\d\phi)\Big\}~,\ \ \  \label{EXACTEXPJOR}
\end{eqnarray}
with $\T^a (q,\delta{q})$ can be identified as the surface term corresponding to the variation of Lagrangian $L$, which is given as 
\begin{eqnarray}
\T^a (q,\delta{q})=\frac{1}{16\pi}[-2g^{ab}\frac{\omega}{\phi}(\nabla_b\phi) \delta\phi +\phi \delta v^a-2(\nabla_b\phi)p^{iabd}\delta g_{id}]~. \label{THETA}
\end{eqnarray}
Here $E_{ab}=0$ and $E_{(\phi)}=0$ corresponds to the equations of motion for the fields $g^{ab}$ and $\phi$, respectively. However, our goal is to obtain the conserved currents (both Noether and ADT currents) off-shell. Therefore, while obtaining those currents, we do not use the equations of motion \textit{i.e.} we nowhere put $E_{ab}=0$ or $E_{(\phi)}=0$.

In the Einstein frame, the arbitrary variation of the Lagrangian yields
\begin{eqnarray}
&&\delta(\sqrt{-\tilde{g}}\tilde{L})=
\sqrt{-\t{g}}\t{E}_{ab}\delta{\t{g}^{ab}}+\sqrt{-\t{g}}\t{E}_{(\t{\phi})}\delta{\t{\phi}}
+\sqrt{-\t{g}}\t{\na}_a\t{\T}^a (\t{q},\delta{\t{q}})~, \label{VAR2}
\end{eqnarray}
where $\t{q}\in \{\t{g}_{ab}, \t{\phi}\}$. The exact expressions of  $\t{E}_{ab}$, $\t{E}_{(\t{\phi})}$ and $\t{\T}^a (\t{q},\delta{\t{q}})$ are given as
\begin{align}
 \t{E}_{ab}=\frac{\tilde{G}_{ab}}{16\pi}-\frac{1}{2}\tilde{\nabla}_a\tilde{\phi}\tilde{\nabla}_b\tilde{\phi}+\frac{1}{4}\tilde{g}_{ab}\tilde{\nabla}^i\tilde{\phi}\tilde{\nabla}_i\tilde{\phi}+\frac{1}{2}\tilde{g}_{ab}U(\tilde{\phi})~;
\no 
\\
\t{E}_{(\t{\phi})}=\t{\na}_a\t{\nabla}^a\tilde{\phi}-\frac{dU}{d\tilde{\phi}}~;\ \ \ \ \ \ \ \ \ \ \ \
\no 
\\
\textrm{and} \ \ \ \ \ \ \ \ \ \ \ \ \ \ \ \ \ \ \ \ \ \ \ \ \
\no 
\\
\t{\T}^a (\t{q},\delta{\t{q}})=\frac{\delta\tilde{v}^a}{16\pi}-(\tilde{\nabla}^a\tilde{\phi})\delta\tilde{\phi}~. \ \ \ \ \ \  \label{EXACTEXEIN}
\end{align}
Again, in this case $\t{E}_{ab}=0$ and $\t{E}_{(\t{\phi})}=0$ corresponds to the equations of motion of the fields $\t g^{ab}$ and $\t\phi$.

Here we clarify that, while obtaining the equation of motion from the action, we have not considered the external matter source. The external matter action in the Jordan frame will be defined as $\mathcal{A}^{(m)}=\int d^4x\mg L^{(m)}$ whereas, the same in the Einstein frame will be given as $\t{\mathcal{A}}^{(m)}=\int d^4x\tmg \t L^{(m)}$. Under the conformal transformation relation \eqref{GAB}, we have
\begin{equation}
\tilde{\mathcal{L}}^{(m)} = \Omega^{-4} \mathcal{L}^{(m)} ~.
\end{equation}
Now, the energy-momentum tensor is defined as
\begin{equation}
\tilde{T}^{(m)}_{ab} = - \frac{2}{\sqrt{-{g}}}\frac{\delta}{\delta g^{cd}} \Big(\sqrt{-g} ~ \mathcal{L}^{(m)}\Big) ~.
\end{equation}
Thus, the energy-momentum tensor in the two frames will be related as
\begin{equation}
\tilde{T}^{(m)}_{~ab} = \Omega^{-2} T^{(m)}_{~ab},~~~~~~~ \tilde{T}^{(m)a}_{~~b} = \Omega^{-4} T^{(m)a}_{~~b},~~~~~~~\tilde{T}^{(m)ab} = \Omega^{-6} T^{(m)ab}~.
\label{conformtab}
\end{equation}
Finally, when the external matter source is present, the equation of motion of the metric tensor will be changed as $E_{ab}=T_{ab}/2$ and $\t E_{ab}=\t T_{ab}/2$ (instead of $E^{ab}=0$ and $\t E^{ab}=0$ respectively).

 In the following we discuss one important relation, which will help us to obtain the conserved Noether current off-shell.

\subsection{Generalized Bianchi identity in scalar-tensor theory of gravity}
In Einstein's GR, the (contracted) Bianchi identity ensures the (local) conservation of energy. In addition, it is also used to obtain the off-shell expression of the Noether current due to the diffeomorphism invariance. Now here we show the corresponding analogue of the Bianchi identity in the ST gravity, which is known as the generalized Bianchi identity or the Noether identity. We provide the expression of generalized Bianchi identity or the Noether identity in each frame

\vskip 1mm    
\noindent
{\bf Jordan frame:}
In the Jordan frame, it can be proved that \cite{Bhattacharya:2018xlq}
\begin{align}
\na_bE^{ab}=-\frac{1}{2}(\na^a\phi) E_{(\phi)}~, \label{RELEE}
\end{align}
which is known as the generalized Bianchi identity of the Jordan frame.
We shall see that, similar to GR, the above expression \eqref{RELEE} will help us to obtain the Noether current off-shell. Note, when external matter is present, the equation of motion of the metric tensor is given by $E^{ab}=8\pi T^{ab}_{(m)}$, where $T^{ab}_{(m)}$ is the energy-momentum tensor of the external matter field. Thus, the above relation \eqref{RELEE} suggests that the (local) conservation of energy-momentum tensor takes place on-shell \textit{i.e.} when $E_{(\phi)}=0$~. This is a stark difference of GR and ST gravity.
\vskip 1mm    
\noindent
{\bf Einstein frame:}
In the Einstein frame, the generalized Bianchi identity is provided as
\begin{align}
\t{\na}_b\t{E}^{ab}=-\frac{1}{2}(\t{\na}^a\t{\phi})\Big[\t{\square}\t{\phi}-\frac{dU}{d\t{\phi}}\Big]=-\frac{1}{2}(\t{\na}^a\t{\phi})\t{E}_{\phi}~. \label{RELEEE}
\end{align}
Again, while obtaining the off-shell Noether current, this generalized Bianchi identity \eqref{RELEEE} will be shown to play a crucial role. In addition, as was the case in the Jordan frame, the local energy conservation takes place on-shell (when $\t{E}_{\phi}=0$).
\subsection{Diffeomorphism invariance and conserved Noether current}
The scalar-tensor gravity is diffeomorphism-invariant theory in both the frames. In section \ref{var}, we have obtained the change in the Lagrangian due to the arbitrary variation. When the change in the Lagrangian is due to the diffeomorphism, $\delta$ is replaced by the Lie-derivative. Thereby, one can obtain the expression of the conserved Noether current and the expression of two-ranked anti-symmetric Noether potential in both the frames. Here, we briefly provide the outline of the procedure to obtain Noether current and potential in each frame.
\vskip 1mm    
\noindent
{\bf Jordan frame:} Due to the diffeomorphism $x^a\longrightarrow x^a+\xi^a$, the change in the Lagrangian can be obtained from \eqref{DELL'} by replacing $\d$ to $\lie$, which yields
\begin{align}
\lie (\sqrt{-g}L')=-2\sqrt{-g}\na_a(E^{ab}\xi_b)+2\sqrt{-g}\xi_b\na_aE^{ab}+\sqrt{-g}E_{(\phi)}\xi^a\na_a\phi+\sqrt{-g}\na_a\T'^a (q,\lie q)~. \label{ptolie}
\end{align}
Now, the LHS of \eqref{ptolie} can be straightforwardly obtained as $\lie (\sqrt{-g}L')=\sqrt{-g}\na_a(L'\xi^a)$ and by using the generalized Bianchi/ Noether identity \eqref{RELEE} on the RHS of \eqref{ptolie}, one obtains $\na_aJ^a=0$, where the conserved off-shell Noether current ($J'^a$) can be identified as
\begin{align}
J'^a=L'\xi^a+2E^{ab}\xi_b-\T'^a (q,\lie q)~. \label{J'A}
\end{align}
The above expression of Noether current can be further written in terms of a two-ranked anti-symmetric Noether potential (\textit{i.e.} $J'^a=\na_bJ'^{ab}$), where the anti-symmetric Noether potential can be obtained as (see \cite{Bhattacharya:2018xlq} for details)
\begin{align}
J'^{ab}=\frac{1}{16\pi}[\nabla^a(\phi\xi^b)-\nabla^b(\phi\xi^a)]~. \label{JAB'}
\end{align}
Note that the above expressions of Noether current and potential, as provided by Eqs \eqref{J'A} and \eqref{JAB'} are obtained from the Lagrangian $L'$ of Eq. \eqref{L'}. For a curious reader, we provide the expressions of Noether current and potential corresponding to the Lagrangian $L$ (as defined in Eq. \eqref{SJ}), which are given as (see derivation in \cite{Bhattacharya:2018xlq})
\begin{align}
J^a=L\xi^a+2E^{ab}\xi_b-\T^a (q,\lie q)~. \label{JAST}
\end{align}
and 
\begin{align}
J^{ab}=\frac{1}{16\pi}\Big[\phi(\nabla^a\xi^b-\nabla^b\xi^a)+2\xi^a(\nabla^b\phi)-2\xi^b(\nabla^a\phi)\Big]~. 
\label{JABJORDAN}
\end{align}
In the following section, when we obtain the 1st law using the Wald's formalism, we show that the thermodynamic parameters obtained from the current $J'^a$ are more appropriate as the thermodynamic parameters, defined by the Noether current $J'^a$ are conformally equivalent. On the contrary, if one defines the thermodynamic parameters in terms of $J^a$, the thermodynamic parameters can be shown to be equivalent only when one incorporates several assumptions, such as the asymptotic flatness of the spacetime \textit{etc.} \cite{Koga:1998un}. Thus, it can be argued that the current $J'^a$ is more appropriate than $J^a$, which also imply that the Lagrangian $L'$ is more appropriate than $L$ in the Jordan frame.
\vskip 1mm    
\noindent
{\bf Einstein frame:} In the Einstein frame, one can similarly obtain the conserved Noether current and potential due to the diffeomorphism invariance. The expression of Noether current is provided as (for details, see \cite{Bhattacharya:2018xlq})
\begin{align}
\t{J}^a=\t{L}\t{\xi}^a+2\t{E}^{ab}\t{\xi}_b-\t{\T}^a (\t{q},\lie \t{q})~, \label{JEINT}
\end{align}
and, the Noether potential is given as
\begin{align}
\t{J}^{ab}=\frac{1}{16\pi}[\tilde{\nabla}^a\tilde{\xi}^b-\tilde{\nabla}^b\tilde{\xi}^a]~. \label{JABEIN}
\end{align}
Thus, we have obtained the Noether current and potential in the two frames. Note, that the obtained Noether currents, in the two frames, are valid for any arbitrary diffeomorphisms ($\xi^a$ and $\t\xi^a$). In the following, we discuss the procedure of obtaining another conserved current which is widely used \textit{i.e.} the ADT current and potential. 

\subsection{The ADT current}
Unlike Noether current due to the diffeomorphiam, the ADT current is not obtained using the symmetry arguments. In Einstein's GR, it can be shown (details follows from the arguments provided in \cite{Abbott:1981ff,Deser:2002rt, Deser:2002jk, Deser:2003vh}) that on-shell (\textit{i.e.} $G_{ab}=0$), using the Bianchi-identity (\textit{i.e. $\nabla_bG^{ab}=0$}), we obtain $\nabla_a\delta G^{ab}=0$, where $\delta G^{ab}$ (known as the linearization tensor) is the {\it{first order}} change in $G^{ab}$ due to the arbitrary perturbation $g_{ab}\longrightarrow g_{ab}+ h_{ab}$. Therefore, $\delta G^{ab}$ is a conserved quantity, albeit a tensor. Since conserved charge is defined in terms of a vector, $\delta G^{ab}$ is contracted with the Killing vector ($\chi$) and the conserved current is defined as $\delta G^{ab}\chi_b$, which is known as the ADT current. Thus, the on-shell conservation of $\delta G^{ab}\chi_b$ requires two major inputs: not only the Killing vector, but also the conservation of $\delta E^{ab}$ (obtained from the Bianchi identity and the equation of motion). The off-shell extension has been formulated by constructing a two-ranked anti-symmetric potential ($J^{ab}_{ADT}$) in such a way that its divergence (\textit{i.e.} $\nabla_b J^{ab}_{ADT}$) is equal to $\delta G^{ab}\chi_b$ added upto terms which are proportional to $G_{ab}$ (for details see \cite{Bouchareb:2007yx}). On-shell, the terms proportional $G_{ab}$ vanish and $\nabla_b J^{ab}_{ADT}$ boils down to $\delta G^{ab}\chi_b$~. Thus, the conservation of $J^{a}_{ADT}$ originates from geometrical arguments and not from the symmetry arguments. Here we have followed the same procedure, and have obtained the expression of the ADT currents in the two frames. For detail mathematical approaches, we refer to our paper \cite{Bhattacharya:2018xlq}.
\vskip 1mm    
\noindent
{\bf Jordan frame:} In the Jordan frame, it can be proved that (for details, see \cite{Bhattacharya:2018xlq})
\begin{align}
J^i_{ADT}|_{on-shell}=\d E^{ij}\chi_j~, \label{JADT}
\end{align} 
is a conserved quantity on-shell. In this case, $\d E^{ij}$ corresponds to the linearized tensor \textit{i.e.} the first order change in $E^{ij}$ due to the perturbation  in the perturbation in the metric $g^{ab}\rightarrow g^{ab}+\d g^{ab}$. The conservation of the $J^i_{ADT}$ follows from the fact that $\na_b\d E^{ab}=0$ on-shell (which can be proved using eq (\ref{RELEE}))  and the property of the Killing vector (i.e. $\na_a\chi_b=-\na_b\chi_a$). Thus, the expression of on-shell ADT current is provided by the Eq. \eqref{JADT}. We call it as the on-shell current as its conservation can be proved using the equation of motion. In the case of off-shell, it can be proved that (for details see \cite{Bhattacharya:2018xlq}):
\begin{align}
\d E^{ij}\chi_j=\na_jJ^{ij}_{ADT}-E^{ik}h_{kj}\chi^j+\frac{1}{2}\chi^iE^{jk}h_{jk}-\frac{1}{2}\chi^jE^i_jh~, \label{DeltaE}
\end{align}
where
\begin{eqnarray}
&&J^{ij}_{ADT}[\chi]=\frac{1}{32\pi}\Big[\phi\Big(\chi^j\na_k h^{ki}-\chi^i\na_k h^{kj}+\chi_k\na^i h^{kj}-\chi_k\na^j h^{ki}+\chi^i(\na^jh)
\no 
\\
&&-\chi^j(\na^ih)+h^{jk}\na_k\chi^i-h^{ik}\na_k\chi^j+h\na^{[i}\chi^{j]}\Big)+(\na_k\phi)\Big(\chi^jh^{ik}-\chi^ih^{jk}\Big)\Big]~.\ \ \ \label{JIJADT}
\end{eqnarray}
and $h_{ab}=\d g_{ab}$ or equivalently $h^{ab}=-\d g^{ab}$. Note that last three terms on the RHS of \eqref{DeltaE} is proportional to $E^{ab}$, which vanish on-shell. Thus, one can identify $J^{ij}_{ADT}$ as the off-shell ADT potential. The off-shell ADT current can be identified as
\begin{align}
J^i_{ADT}|_{off-shell}=\d E^{ij}\chi_j+E^{ik}h_{kj}\chi^j-\frac{1}{2}\chi^iE^{jk}h_{jk}+\frac{1}{2}\chi^jE^i_jh~. \label{JADTOFF}
\end{align}
Now that we have obtained the ADT current and potential in the Jordan frame, we now move on to obtain the corresponding counterpart in the Einstein frame.

\vskip 1mm    
\noindent
{\bf Einstein frame:} As was the case for the Jordan frame, off-shell $\d \t E^{ij}\t\chi_j$ can be written as \cite{Bhattacharya:2018xlq}, 
\begin{align}
\d \t E^{ij}\t\chi_j=\t\na_j\t J^{ij}_{ADT}-\t E^{ik}\t h_{kj}\t\chi^j+\frac{1}{2}\t \chi^i\t E^{jk}\t h_{jk}-\frac{1}{2}\t \chi^j\t E^i_j\t h~, \label{DeltaE2}
\end{align}
where, the two-ranked anti-symmetric ADT potential $\t J^{ij}_{ADT}$ can be obtained as
\begin{align}
\t J^{ij}_{ADT}[\t\chi]=\frac{1}{32\pi}\Big[\t \chi^j\t\na_k \t h^{ki}-\t \chi^i\t \na_k \t h^{kj}+\t\chi_k\t\na^i \t h^{kj}-\t \chi_k\t\na^j \t h^{ki}
\no 
\\
+\t\chi^i(\t\na^j\t h)-\t\chi^j(\t\na^i\t h)+\t h^{jk}\t\na_k\t\chi^i-\t h^{ik}\t\na_k\t\chi^j+\t h\t\na^{[i}\t\chi^{j]}\Big]~, \label{JIJADTTIL}
\end{align}
where  $\t h_{ab}=\d \t g_{ab}$, $\t h^{ab}=-\d \t g^{ab}$. Thus the off-shell ADT current can be identified as

\begin{align}
\t J^i_{ADT}|_{off-shell}=\d \t E^{ij}\t\chi_j+\t E^{ik}\t h_{kj}\t\chi^j-\frac{1}{2}\t \chi^i\t E^{jk}\t h_{jk}+\frac{1}{2}\t \chi^j\t E^i_j\t h~. \label{JADTOFFTIL}
\end{align}
On-shell, $\t E^a_b=0$. Therefore, the on-shell expression of ADT current in the Einstein frame is given by the first term on the RHS of Eq. \eqref{JADTOFFTIL} \textit{i.e.} $\t J^i_{ADT}|_{on-shell}=\d \t E^{ij}\t\chi_j$. 
As it is apparent from the analysis, the ADT currents and the potentials, in the two frames, are defined for the Killing vectors ($\chi^a$ and $\t\chi^a$) only.\footnote{Recently we have obtained a generalised ADT-like current, which is obtained for arbitrary diffeomorphism in GR and Lanczos-Lovelock gravity \cite{Bhattacharya:2018tse}.}

 Again, we emphasize that the ADT current (either on-shell or off-shell) has not been obtained using the symmetry argument \textit{i.e.}, identifying those symmetries which leave the action invariant (or change upto a total derivative) and obtaining conserved current using the Noether's theorem. Instead, a quantity has been defined under linear perturbation from the equation of motion and then contracted with the timelike Killing vector (assuming it exists for the spacetime under study). This process yields a first rank tensor. Interestingly, it has been observed that such one is covariantly conserved and it is called ADT current. It shows that ADT current does not have origin through Noether prescription due to a particular symmetry of spacetime. Moreover in the original approach \cite{Abbott:1981ff,Deser:2002rt, Deser:2002jk, Deser:2003vh} the construction of it and its covariant conservation greatly depend upon the use of equation of motion for gravity. In that sense such is called as on-shell ADT current.

On the other hand Noether current for gravity  has been obtained through Noether prescription,  arising due to the diffeomorphism invariance of the theory. Now diffeomorphism is a local gauge symmetry and hence the corresponding current is off-shell conserved. In literature the nomenclature, off-shell, is usually reserved for this kind of analysis. In the case of ADT current, it has been observed that one can construct a first rank tensor which is covariantly conserved even without use of equation of motion (see \cite{Bouchareb:2007yx} for details). Therefore this new construction does not depend on the use of equation of motion of gravity and, in this sense, the new construction is called as off-shell ADT current in literature (e.g. see \cite{Bouchareb:2007yx}). Here we also followed the same. However, as the Noether current and ADT current are different by construction, they can not be treated in the same footing. Therefore, the word off-shell has been used for ADT current in a different sense, unlike what it is used for conserved Noether current.

\subsection{Connection between Noether and ADT current}
Although the Noether current and the ADT current formalism has been developed independently, it can be shown that they are connected to each other. As a matter of which, it can be argued that the two covariant formalisms to obtain the first law \textit{i.e.} the Wald's approach (that incorporates the Noether current) and the ADT approach (which incorporates the ADT current) are equivalent. In the following, we show the connection of the Noether and the ADT current in each frame.
\vskip 1mm    
\noindent
{\bf Jordan frame:} In the Jordan frame, the Noether and the ADT currents are related as \cite{Bhattacharya:2018xlq}
\begin{align}
\sqrt{-g}J^{ij}_{ADT}[\chi]=\frac{1}{2}\d(\sqrt{-g}J'^{ij}[\chi])-\sqrt{-g}\chi^{[i}\T'^{j]}(q,\d q)~. \label{NOADT}
\end{align}
In \eqref{NOADT}, $J'^{ij}[\chi]$ is the Noether potential corresponding to the Killing diffeomorphism and not of any arbitrary diffeomorphism. On the other hand, ADT current is defined only for the Killing vectors. Thus, we can say, the above relation \eqref{NOADT} expresses the connection of the ADT current with the conserved Noether current corresponding to Killing diffeomorphism.
\vskip 1mm    
\noindent
{\bf Einstein frame:} In the Einstein frame, the relation between the Noether current (corresponding to the Killing diffeomorphism) and the ADT current is provided as \cite{Bhattacharya:2018xlq}
\begin{align}
\sqrt{-\t g}\t J^{ij}_{ADT}[\t\chi]=\frac{1}{2}\d(\sqrt{-\t g}\t J^{ij}[\t\chi])-\sqrt{-\t g}\t \chi^{[i}\t \T^{j]}(\t q,\d \t q)~. \label{NOADT'}
\end{align}

Now that we have obtained the expressions of the conserved currents and have shown the connection of the Noether and the ADT currents, we have set the stage nicely to obtain the thermodynamic laws in a covariant way. In the following section, we obtain the thermodynamic laws in the two frames.

%%%%%%%%%%%%%%%%%%%%%%%%%%%%%%%%%%%%%%%%%%%%%%%%%%%%%%%%%%%%%%%%%%%%%%%%%%%%%%%%%%%
\section{Thermodynamics in scalar-tensor gravity}
Black hole thermodynamics has been one of the key aspects of general relativity. Incidentally this thermodynamic concept can be extended on any null surface (e.g. Killing horizon) which provides the idea of the thermodynamic structure of gravitational theories. In the absence of a proper quantum theory of gravity, such thermodynamic interpretaion provides an alternative viewpoint \textit{i.e.} gravity can be interpreted as an emergent phenomena and not any fundamental force. Several studies \cite{Eling:2006aw,Elizalde:2008pv,Chirco:2009dc,Padmanabhan:2002sha,Padmanabhan:2003gd,Faraoni:2010yi,Bamba:2009gq} suggest that the thermodynamic aspect of GR should transcend well-beyond GR and can be found in modified theories of gravity as well. In the context of scalar-tensor theory (or Brans-Dicke theory), it has been found that the Bekenstein-Hawking area law of entropy does not hold. Instead, the black hole entropy in this theory is proportional to the area of the black hole horizon as well as the scalar field $\phi$ \cite{Kang:1996rj, Jacobson:1993vj,Iyer:1994ys,Visser:1993nu}. However, a systematic formulation of black hole  thermodynamics in scalar-tensor gravity has not been obtained earlier until recently \cite{Bhattacharya:2017pqc,Bhattacharya:2018xlq,Bhattacharya:2020jgk}. In one work of Koga and Maeda \cite{Koga:1998un}, the thermodynamic laws were obtained only after incorporating certain assumptions. These demonstrate that a proper formulation of thermodynamics in ST gravity has always been challenging. As we discuss later, several non-trivial issues arise while formulating the thermodynamic laws in scalar tensor theory, which are not there in GR. In the following, we briefly discuss the problems which one encounters during the formulation of thermodynamics in scalar-tensor gravity. Then we move on to discuss how those problems can be overcome and the thermodynamic laws can be established in the two frames. Moreover, it will be shown that the thermodynamic parameters in the two frames are equivalent in the two frames. 
%%%%%%%%%%%%%%%%%%%%%%%%%%%%%%%%%%%%%%%%%%%%%%%%%%%%%%%%%%%%%%%%%%%%%%%%%%%%%%%%%%
\subsection{Challenges to obtain thermodynamic description in scalar-tensor gravity}
The main issue while formulating thermodynamic descriptions in the two frames stems from the fact that the ST gravity is described in the two frames which are conformally connected. As a matter of which, not only one has to formulate the thermodynamic laws, but also one has to worry about whether the defined thermodynamic parameters are equivalent in the two frames. Moreover, from the work of Kang \cite{Kang:1996rj}, it was found that the black hole entropy in the Jordan frame is given as $S=\phi A/4$, where $A$ is the area of the BH horizon (\textit{i.e.} $A=\int_{\mathcal{H}}\sqrt{\sigma} d^2x$, where $\sigma$ is the determinant of the induced metric on the BH horizon).  However, for a long time, it is known that the BH entropy is conformally invariant. The null geodesics are left unchanged due to the conformal transformation. Therefore, a black hole, which is also a null surface, remains unchanged under the conformal rescaling. On the contrary, the area/length changes under the conformal transformation. In Einstein frame, the entropy ($\t S$) is given only by the area ($\t A$) of the BH horizon. However, one can obtain $\t A=\int_{\mathcal{H}}\sqrt{\t\sigma} d^2x$=$\int_{\mathcal{H}}\phi\sqrt{\sigma} d^2x=\phi A$~, which imply that the entropy in the two frames are equivalent. Although we have this straightforward proof regarding the equivalence of the black hole entropy, major problem arises in defining the thermodynamic energy. In literature, there are several prescriptions of energy, most of which are not conformally invariant (such as the Misner-Sharp energy, Hawking-Heyward quasilocal energy \textit{etc.}). Thus, it becomes challenging to find a suitable candidate of energy which can play the role of internal energy in ST gravity.

In the following, we obtain the thermodynamic relations of the two frames in a covariant way and, thereby, we show that all ther thermodynamic parameters are equivalent. Firstly, we perform our analysis for the Killing horizon as it is relatively simpler. Then we obtain the first law for a generic null surface.
%%%%%%%%%%%%%%%%%%%%%%%%%%%%%%%%%%%%%%%%%%%%%%%%%%%%%%%%%%%%%%%%%%%%%%%%%%%%%%%
\subsection{Thermodynamic description for the Killing horizon}
In black hole thermodynamics, the Killing horizon has played a pivotal role. As it was shown by Hawking (in the context of GR), the event horizon of a static spacetime is a Killing horizon \cite{Hawking:1971vc}. In addition, assumption of the Killing symmetry in the spacetime largely simplifies the analysis. Therefore, for simplicity, we start the analysis for the Killing horizon. Moreover, here we adopt the Wald's formulation to obtain the first law, which is a powerful way to obtain the first law and to define the thermodynamic parameters in a covariant manner. Another reason to consider the Killing horizon at first is that Wald's formulation is defined for the Killing horizon only. Thus, here we consider that the black hole horizon is a Killing horizon in \textit{both the frames}. Before moving on, we mention one important comment. Note that the rigidity theorem \cite{Hawking:1971vc}, which ensures that the event horizon of a stationary black hole is also a Killing horizon, has not been proved in the context of the scalar-tensor theory yet and it is an important subject to work in future. Therefore, we claim that our analysis is valid only for the Killing horizon. Since the proof of the rigidity theorem is still missing for scalar-tensor theory, it is not sure whether it will be valid for an event horizon in scalar-tensor gravity.

Firstly, we consider a Killing vector $\tilde{\boldsymbol{\chi}}$ in the Einstein frame $(\mathcal{M}, \tilde{g}_{ab}, \tilde{\phi})$~, which is also the generator of the Killing horizon $\tilde{\mathcal{H}}^{(K)}$ (in Einstein frame). Thus, from the Killing vector condition, we have
\begin{equation}
\pounds_{\tilde{\boldsymbol{\chi}}} ~\tilde{g}_{ab} \overset{(\mathcal{M},\tilde{\boldsymbol{g}},\tilde{\phi})}{=} 0 ~,
\label{disc}
\end{equation}  
Using the conformal transformation relation \eqref{GAB}, we arrive at the Jordan frame $(\mathcal{M}, g_{ab}, \phi)$ and we obtain
\begin{equation}
\pounds_{\tilde{\boldsymbol{\chi}}}~g_{ab} \overset{(\mathcal{M}, \bf{g}, \phi)}{=} - \frac{1}{\phi} \Big(\pounds_{\tilde{\boldsymbol{\chi}}}~\phi \Big) g_{ab} ~.
\end{equation}
Thus, we find that $\tilde{\boldsymbol{\chi}}$ is a conformal Killing vector unless $\pounds_{\tilde{\boldsymbol{\chi}}}~ \phi  \overset{(\mathcal{M}, \bf{g}, \phi)}{=} 0$. Therefore, we impose the condition
\begin{equation}
\pounds_{\tilde{\boldsymbol{\chi}}}~ \phi  \overset{(\mathcal{M}, \bf{g}, \phi)}{=} 0 ~,
\label{constraint}
\end{equation}
Using Eq. \eqref{PHI}, one can obtain the analogous condition in Einstein frame as
\begin{equation}
\pounds_{\tilde{\boldsymbol{\chi}}}~ \t\phi  \overset{(\mathcal{M}, \bf{\t g}, \t\phi)}{=} 0 ~.
\label{constrainte}
\end{equation}
Now, according to our notational convention, we define the Killing vector, which is the generator of the Killing horizon $\mathcal{H}^{(K)}$ (in the Jordan frame) is $\boldsymbol{\chi}$~. Thus, we have 
\begin{equation}
\tilde{\chi}^a \overset{(\mathcal{M}, g_{ab}, \phi)}{=} \chi^a ~. \label{CHICHITIL}
\end{equation}
Although the above expression shows that the contravariant components of the Killing vectors are the same (under the assumption \eqref{constraint} or equivalently \eqref{constrainte}), the covariant components of the Killing vectors in the two frames are not exactly the same, they are related by the conformal factor due to the transformation relation \eqref{GAB}. Also note that the relation \eqref{CHICHITIL} (which is consequence of the imposed condition \eqref{constraint} or \eqref{constrainte}) are followed from \cite{Jacobson:1993pf} and has also been imposed in \cite{Koga:1998un} in the context of ST gravity. Due to these imposed condition, the Killing horizon in one frame (say $\mathcal{H}^{(K)}$) maps to the Killing horizon of the other frame ($\tilde{\mathcal{H}}^{(K)}$). With these preliminary discussions about the Killing horizons, we obtain the thermodynamic laws for  $\mathcal{H}^{(K)}$ and $\tilde{\mathcal{H}}^{(K)}$ in the following.

\subsubsection{The zeroth law} 
As we have discussed earlier, the Killing vectors ($\boldsymbol{\chi}$ and $\tilde{\boldsymbol{\chi}}$) are the generators of the of the Killing horizons ($\mathcal{H}^{(K)}$ and $\tilde{\mathcal{H}}^{(K)}$ respectively), which becomes null on the horizon. Therefore, on the horizon, the Killing vectors also satisfy the null geodesic conditions, which are given as
\begin{equation}
\tilde{\chi}^b \tilde{\nabla}_b \tilde{\chi}^a \overset{\tilde{\mathcal{H}}^{(K)}}{=} \tilde{\kappa} \tilde{\chi}^a
\label{geodesice}
\end{equation}
and
\begin{equation}
\chi^b \nabla_b \chi^a \overset{\mathcal{H}^{(K)}}{=} \kappa \chi^a ~.
\label{geodesicj}
\end{equation}
Here $\kappa$ and $\t\kappa$ are the non-affinity parameters of the null surface, which can also be identified as the surface gravity of the Killing horizon. Under the condition \eqref{constraint} (or equivalently \eqref{constrainte}), we obtain $\kappa$ and $\t\kappa$ are equivalent \textit{i.e.} $\t\kappa=\kappa$ \cite{Jacobson:1993pf}. 

Here, we shall prove that $\kappa$ (or $\t\kappa$) is constant on the horizon $\mathcal{H}^{(K)}$ (or $\t{\mathcal{H}}^{(K)}$ in the Einstein frame). Later, when we prove the first law, we shall see that the same $\kappa$ (or $\t\kappa$) appears as the conjugate term of the black hole entropy. This will provide the justification for identifying $\kappa/2\pi$ (or $\t\kappa/2\pi$) as the black hole temperature in the two frames. Although, while proving the constancy of $\kappa$ (or $\t\kappa$), we shall use the well-defined formalisms Riemannian manifold, the earlier approaches (which are formulated in the context of GR), does not guarantee the constancy of $\kappa$ (or $\t\kappa$) as the proof of the zeroth law involves the dynamical equations of ST gravity in the two frames. In the following, we prove the zeroth law in the two frames. The method which we adopt here has been followed from \cite{Wald:1984rg}. The proof of the zeroth law, in this approach, relies on two crucial considerations:
\begin{enumerate}
\item \textit{Null dominant energy condition:} It states that, in addition to the weak energy condition ($T^{(m)}_{ab}k^ak^b\geq 0$, where $k^a$ is a future pointing null vector and $T^{(m)}_{ab}$ corresponds to the energy-momentum tensor of external matter field), $-T^{(m)}_{ab}k^a$ (which, roughly, corresponds to the momentum measured by the observer) must be a future pointing causal vector field (\textit{i.e.} either timelike or null) as the flow of mass-energy cannot be faster than the speed of light.
\item \textit{Topology of the horizon being $\mathbb{R} \times \mathcal{J}$:} It implies that the cross-section of the black hole horizon is a spacelike compact two-surface ($\mathcal{J}$). Thus, the tangent space of the horizon is either null or spacelike.
\end{enumerate}

\vskip 1mm    
\noindent
{\bf Jordan frame:} In order to show that the surface gravity ($\kappa$) is unchanged on the horizon, we need to define the directional derivative which lies on the tangent plane of $\mathcal{H}^{(K)}$. It is found that $\epsilon^{abcd}\chi_a$ lies on the tangent plane of $\mathcal{H}^{(K)}$ as $\epsilon^{abcd}\chi_a\chi_b=0$, where $\epsilon^{abcd}$ is the spacetime volume form. Therefore, one can argue that $\epsilon^{abcd}\chi_a\na_b$ is the derivative on the tangent plane of $\mathcal{H}^{(K)}$. Accounting the anti-symmetric property of $\epsilon^{abcd}$, the defined directional derivative $\epsilon^{abcd}\chi_a\na_b$ can be equivalently written as $\chi_{[a} \nabla_{b]}$~. Therefore, our goal is proving $\chi_{[a} \nabla_{b]}\kappa=0$~. However, on the Killing horizon $\mathcal{H}^{(K)}$ it can be proved \cite{Wald:1984rg}
\begin{equation}
\chi_{[a} \nabla_{b]} \kappa \overset{\mathcal{H}^{(K)}}{=} - \chi_{[a} R_{b]}^{~~f} \chi_f ~,
\label{waldonkilling}
\end{equation}
where $R_{ab}$ is the Ricci tensor. Thus, our final goal boils down to prove $\chi_{[a} R_{b]}^{~~f} \chi_f=0$ on the horizon, which will prove the constancy of $\kappa$ on $\mathcal{H}^{(K)}$~. Using the equation of motion $E_{ab}=T_{ab}^{(m)}/2$, where $E_{ab}$ is defined in \eqref{EXACTEXPJOR}, one can obtain
\begin{eqnarray}
-\chi_{[a} R_{b]}^{~~f} \chi_f &~\overset{(\mathcal{M}, g_{ab}, \phi)}{=}~ -\frac{1}{\phi} \Big( 8 \pi \chi_{[a} T_{b]}^{(m)f} \chi_f - \frac{\omega}{2 \phi} \chi_{[a} \delta_{b]}^{~f}\chi_f~(\nabla_i \phi \nabla^i \phi) + \frac{\omega}{\phi} \chi_{[a} \nabla_{b]} \phi ~ (\chi^f \nabla_f \phi) \nonumber 
\\
& - \frac{1}{2} \chi_{[a} \delta_{b]}^{~f} \chi_f ~ V(\phi) + (\chi_{[a} \nabla_{b]}\nabla^{f} \phi) \chi_f - \chi_{[a} \delta_{b]}^{~f} \chi_f (\nabla_i \nabla^i \phi)\Big) \nonumber
\\
& - \frac{1}{2} \chi_{[a} \delta_{b]}^{~f} \chi_f ~R ~.
\end{eqnarray}
Using the imposed condition in the Jordan frame (\ref{constraint}) the above result can be further simplified as
\begin{equation}
\chi_{[a} R_{b]}^{~~f} \chi_f{=}~ \frac{1}{ \phi} \Big(8 \pi \chi_{[a} T^{(m)f}_{b]} \chi_f + \chi^f (\chi_{[a}\nabla_{b]}\nabla_f ~\phi)\Big) ~.
\label{zeroth1j}
\end{equation}
Using Frobenius' hypersurface orthogonality condition (\textit{i.e.} $\chi_{[a}\na_b\chi_{c]}=0$) and using $\liek(\na_a\phi)=0$ (which can be proved straightforwardly from the condition \eqref{constraint}), one can prove that the last term of \eqref{zeroth1j} vanishes on the horizon $\mathcal{H}^{(K)}$ and, thereby, \eqref{zeroth1j} yields
\begin{align}
\chi_{[a} R_{b]}^{~~f} \chi_f\overset{\mathcal{H}^{(K)}}{=}~ \frac{1}{ \phi} 8 \pi \chi_{[a} T^{(m)f}_{b]} \chi_f~. \label{penulzero}
\end{align}
We now use another geometric condition: null Raychaudhuri equation (NRE). For the Killing horizon, the expansion scalar, shear tensor and the deformation tensor -- all vanish. Therefore, the NRC on $\mathcal{H}^{(K)}$ yields
\begin{equation}
R_{ab} \chi^a \chi^b ~\overset{\mathcal{H}^{(K)}}{=}~ 0 ~.
\label{Rabchiachibj}
\end{equation}
In the above relation \eqref{Rabchiachibj}, one can replace $R_{ab}$ using the equation of motion $E_{ab}=T_{ab}^{(m)}/2$, where $E_{ab}$ is defined in \eqref{EXACTEXPJOR}. This imply
\begin{equation}
T^{(m)}_{ab} \chi^a \chi^b ~\overset{\mathcal{H}^{(K)}}{=}~ 0 ~,
\label{Tabchiachibj}
\end{equation}
as all the terms containing $\phi$ vanish due to the condition \eqref{constraint} and also by using $\liek(\na_a\phi)=0$. Since $\chi^b$ is normal to $\mathcal{H}^{(K)}$, it implies that $T^{(m)}_{ab} \chi^a$ lies on the tangent plane of the Killing horizon $\mathcal{H}^{(K)}$, which means $T^{(m)}_{ab} \chi^a$ is either null or spacelike. On the other hand, we have accounted the null dominant energy condition, which imply $T^{(m)}_{ab} \chi^a$ can be either timelike or null. Since the option of being a timelike vector has been ruled out earlier, the only option one left with is: $T^{(m)}_{ab} \chi^a$ is null on $\mathcal{H}^{(K)}$ and, hence, collinear to $\chi_b$. Thus, one can obtain
\begin{equation}
T^{(m)a}_{~~b} \chi^b ~\overset{\mathcal{H}^{(K)}}{=}~ \alpha \chi^a ~, \label{0setup}
\end{equation}
where $\alpha$ is the proportionality factor. The straightforward use of Eq. \eqref{0setup} in \eqref{penulzero}, yields $\chi_{[a} R_{b]}^{~~f} \chi_f=0$, which implies the constancy of $\kappa$ on the Killing horizon $\mathcal{H}^{(K)}$ and, thereby, the zeroth law is proven in the Jordan frame.

\vskip 1mm    
\noindent
{\bf Einstein frame:} In the Einstein frame, we can prove the constancy of $\t\kappa$ following the similar arguments as of the Jordan frame. In this frame, we have the following geometric identity
\begin{equation}
\tilde{\chi}_{[a} \tilde{\nabla}_{b]} \tilde{\kappa}  \overset{\tilde{\mathcal{H}}^{(K)}}{=} - \tilde{\chi}_{[a} \tilde{R}_{b]}^{~~f} \tilde{\chi}_f ~.
\label{waldonkillinge}
\end{equation}
As the arguments provided in the Jordan frame, proving the condition $\tilde{\chi}_{[a} \tilde{R}_{b]}^{~~f} \tilde{\chi}_f$ will prove the constancy of $\t\kappa$ on $\t{\mathcal{H}}^{(K)}$~. Using the field equation \eqref{EXACTEXEIN}, we obtain
\begin{eqnarray}
\tilde{\chi}_{[a} \tilde{R}_{b]}^{~~f} \tilde{\chi}_f &= 16 \pi \Big[\frac{1}{2} \tilde{\chi}_{[a}\tilde{T}^{(m) f}_{~b]} \tilde{\chi}_f + \frac{1}{2} \tilde{\chi}_{[a} \tilde{\nabla}_{b]} \tilde{\phi}~(\tilde{\nabla}^{f} \tilde{\phi} \tilde{\chi}_f) -\frac{1}{4} \tilde{\chi}_{[a} \delta_{b]}^{~f} \tilde{\chi}_f ~(\tilde{\nabla}^i \tilde{\phi} \tilde{\nabla}_i\tilde{\phi}) \nonumber
\\
&-\frac{1}{2}\tilde{\chi}_{[a} \delta_{b]}^{~f} \tilde{\chi}_f ~U(\tilde{\phi}) + \frac{1}{32 \pi} \tilde{\chi}_{[a} \delta_{b]}^{~f} \tilde{\chi}_f ~ \tilde{R}\Big] ~.
\end{eqnarray}
Using the condition imposed on $\t\phi$ \eqref{constrainte}, one can simplify the above equation as
\begin{equation}
 \tilde{\chi}_{[a} \tilde{R}_{b]}^{~~f} \tilde{\chi}_f \overset{\tilde{\mathcal{H}}^{(K)}}{=} 8 \pi \tilde{\chi}_{[a} \tilde{T}^{(m) f}_{b]} \tilde{\chi}_f ~.
\label{zeroth1e}
\end{equation}
The null Raychaudhuri equation (NRE) on the horizon $\t{\mathcal{H}}^{(K)}$ yields 
\begin{equation}
\tilde{R}_{ab}\tilde{\chi}^a\tilde{\chi}^b ~\overset{\tilde{\mathcal{H}}^{(K)}}{=}~ 0 ~,
\label{Rabchiachibe}
\end{equation}
as, for the Killing horizon, the expansion scalar and the shear tensor vanish. Now, in \eqref{Rabchiachibe}, we replace $\tilde{R}_{ab}\tilde{\chi}^a\tilde{\chi}^b$ using the field equation \eqref{EXACTEXEIN} (along with \eqref{constrainte}) and, thereby, we obtain
\begin{equation}
\tilde{T}^{(m)}_{~ab}\tilde{\chi}^a\tilde{\chi}^b ~\overset{\tilde{\mathcal{H}}^{(K)}}{=}~ 0 ~.
\label{nece}
\end{equation}
Since $\t\chi^a$ is normal to $\t{\mathcal{H}}^{(K)}$, the above relation \eqref{nece} imply that $\tilde{T}^{(m)}_{~ab}\tilde{\chi}^a$ lies on the tangent plane of the Killing horizon $\t{\mathcal{H}}^{(K)}$. Therefore, it can be concluded that $\tilde{T}^{(m)}_{~ab}\tilde{\chi}^a$ is either normal or spacelike tensor. Now, we assume that the matter source obeys the null dominant energy condition (NDEC) in the Einstein frame, which imply that $\tilde{T}^{(m)}_{~ab}\tilde{\chi}^a$ has to be causal (\textit{i.e.} either timelike or null). Since $\tilde{T}^{(m)}_{~ab}\tilde{\chi}^a$ cannot be timelike as argued above, the only option left is: $\tilde{T}^{(m)}_{~ab}\tilde{\chi}^a$ is null on the horizon(\textit{i.e.} collinear to the Killing vector on the horizon). Thus, we have
\begin{equation}
\tilde{T}^{(m) a}_{~~~~~~b} \tilde{~\chi}^b ~\overset{\tilde{\mathcal{H}}^{(K)}}{=}~ \tilde{\alpha} \tilde{\chi}^a ~,
\label{NDECconsequencee}
\end{equation} 
where $\tilde{\alpha}$ is the proportionality factor. Using \eqref{NDECconsequencee} in \eqref{zeroth1e}, we finally obtain 
\begin{equation}
\tilde{\chi}_{[a} \tilde{\nabla}_{b]} \tilde{\kappa}  \overset{\tilde{\mathcal{H}}^{(K)}}{=} 0 ~,
\label{ZEROEIN}
\end{equation}
which implies the constancy of $\kappa$ on the horizon $\tilde{\mathcal{H}}^{(K)}$ and, hence, the zeroth law is proven in the Einstein frame.

We end up the discussion on the zeroth law with the following comments. From the discussion presented here, it might seem to the reader that the null dominant energy condition has been imposed twice as we have individually accounted NDEC in the two frames. However, it is not the case. Since the energy momentum tensor in the two frames are connected to each other by the relation \eqref{conformtab}, the NDEC in one frame corresponds to the NDEC of the other.

\subsubsection{First law on the Killing horizon}
In the earlier section, we have established the constancy of $\kappa$ (or $\tilde{\kappa}$) on the Killing horizon. In this section, we shall establish the first law in a covariant manner. Firstly we obtain the first law on the Killing horizon. Then, we generalize the first law for a generic null surface. Moreover, we shall show that the thermodynamic parameters, which are defined in these approaches are conformally invariant. In addition, this will also allow us to interpret $\kappa$ and $\t\kappa$ as the temperatures in the two frames. The proof of the first law using the Wald's formalism depends crucially on two important considerations: (i) the Killing symmetry of the spacetime, (ii) presence of a bifurcation Killing horizon (which is two-dimensional spacelike cross-section of two Killing horizons generated by the same Killing vector $\chi^a$; as a result, $\chi^a=0$ on the bifurcation surface). The procedure of obtaining the first law, following the Wald's formalism, are discussed as follows.
\vskip 1mm    
\noindent
{\bf Jordan frame:} The proof of the first law using the Wald's formalism is provided using the conserved Noether current, which we have obtained earlier. Here we take an arbitrary variation (due to the change in the dynamic fields $g_{ab}$ and $\phi$) of the conserved Noether current on the Killing horizon with the on-shell condition and we show that this variation leads to the first law. The Noether current in the Jordan frame is defined in \eqref{J'A}. The variation of Noether current corresponding to the Killing diffeomorphism $J'^a[\chi]$ (as our aim is to obtain the thermodynamic law on the Killing horizon) yields
\begin{align}
\delta(\sqrt{-g}J'^a[\chi])=\d(\sqrt{-g}L')\chi^a-\d [\sqrt{-g}\T'^a(q, \liek q)]~. \label{DELJ}
\end{align}
Here $\delta\chi^a=0$ as, in this case, $\d$ represents the variation due to the change in the variables $q\in \{g_{ab}, \phi\}$. Using Eq. \eqref{DELL'} (along with the on-shell conditions \textit{i.e.} $E_{ab}=0$ and $E_{(\phi)}=0$) it yields
\begin{align}
\delta(\sqrt{-g}J'^a[\chi])=\sqrt{-g}[\na_i\T'^i(q, \d q)]\chi^a-\d [\sqrt{-g}\T'^a(q, \liek q)]~. \label{VAR}
\end{align}
Now, from a straightforward computation one can obtain the following identity
\begin{align}
\pounds_{\chi}[\sqrt{-g} \T'^a(q, \d q)]=\sqrt{-g}\chi^a\nabla_i[\T'^i(q, \d q)]-2\sqrt{-g}\nabla_b[\chi^{[a}\T'^{b]}(q, \d q)]~,
\end{align}
where $A^{[a}B^{b]}=(1/2)(A^aB^b-A^bB^a)$. Using the above identity in \eqref{VAR}, one can obtain
\begin{align}
\delta(\sqrt{-g}J'^a[\chi])=\pounds_{\chi}[\sqrt{-g} \T'^a(q, \d q)]-\d [\sqrt{-g}\T'^a(q, \liek q)]+2\sqrt{-g}\nabla_b[\chi^{[a}\T'^{b]}(q, \d q)]~. \label{VARJST2}
\end{align}
One can now identify 
\begin{align}
\o^a=-\pounds_{\chi}[\sqrt{-g} \T'^a(q, \d q)]+\d [\sqrt{-g}\T'^a(q, \liek q)]~, \label{O1}
\end{align} 
as the variation of the symplectic Hamiltonian density with the following arguments. Remember, in classical mechanics, the variation of the Lagrangian $L(x_i, \dot{x_i})$ is given as $\delta L(x_i, \dot{x_i})=[(\frac{\partial L}{\partial x_i})-d_t(\frac{\partial L}{\partial \dot{x_i}})]\delta x_i +d_t[p^i\delta x_i]$ where $x_i$ corresponds to the generalized coordinate and $p^i=\frac{\partial L}{\partial \dot{x_i}}$ corresponds to the generalized momentum. The Euler-Lagrange's equation of motion vanishes on-shell and, the variation of the Hamiltonian ($H(x_i,p^i)=p^i(d_tx_i)-L(x_i, \dot{x_i})$) under the arbitrary variation of $x_i$ is provided as 
\begin{align}
\delta H(x_i,p^i)=\delta[p^i(d_tx_i)]-d_t[p^i(\delta x_i)]~. \label{DELHST2}
\end{align}
Comparing \eqref{O1} and \eqref{DELHST2} and with the following analogous correspondence
$\sqrt{-g} \T'^a(q, \d q)\equiv p^i(\delta x_i)$ and $\sqrt{-g}\T'^a(q, \lie q)\equiv p^i(d_tx_i)$, one can argue $\o^a$ as the variation of the symplectic Hamiltonian density. Now, with the identification \eqref{O1}, Eq. \eqref{VARJST2} yields
\begin{align}
\omega^a=-\delta(\sqrt{-g}J'^a[\chi])+2\sqrt{-g}\nabla_b[\chi^{[a}\T'^{b]}(q, \d q)].  \label{O2}
\end{align}
Thus, the total change in the Hamiltonian is provided as
\begin{eqnarray}
&&\delta H[\chi]= \int_c d\Sigma_a \frac{\omega^a}{\sqrt{-g}}=-\d\int_cd\Sigma_a\nabla_b(J'^{ab} [\chi])+2\int_c d\Sigma_a\nabla_b[\chi^{[a}\T'^{b]}(q, \d q)]~, 
\label{DELH1}
\end{eqnarray} 
where, $c$ symbolizes the Cauchy hypersurface, upon which the integration is been performed. In addition the elemental surface area of the three-dimensional Cauchy hypersurface has been defined as $d\Sigma_a=n_a\sqrt{h}d^3x$ where $n_a$ is the normal and $h$ is the determinant of the induced metric defined on $c$. Since, each term on the RHS of \eqref{DELH1} is a total derivative term, one can apply Stoke's law and, thereby, the three integrations can be written as two-surface integration. Now, this two-surface is not a compact one, and it has two edges. The inner surface of $c$ is considered as the bifurcation surface {\it i.e} $\mathcal{H}$ which is the cross-section of the Killing horizon $\mathcal{H}^{(K)}$. As a result, $\chi^a=0$ on $\mathcal{H}$. One the other hand, the outer two-surface lies on the asymptotic infinity ({\it i.e} $\p c_{\infty}$). 
Therefore, from \eqref{DELH1} one can obtain,
\begin{align}
\delta H[\chi]=-\frac{1}{2}\delta\int_{\mathcal{H}} d\Sigma_{ab}J'^{ab}[\chi]+\frac{1}{2}\delta\int_{\partial c_{\infty}} d\Sigma_{ab}J'^{ab}[\chi]-\int_{\partial c_{\infty}} d\Sigma_{ab} \chi^{[a}\T'^{b]}(q, \d q)~. \label{DELH2}
\end{align}
Since $\chi^a=0$ on $\mathcal{H}$, the contribution coming from the term $\chi^{[a}\T'^{b]}(q, \d q)$ vanishes on $\mathcal{H}$.
 Moreover, it can be proved that the symplectic Hamiltonian (as defined in \eqref{O1}) corresponding to a Killing vector always vanish. Therefore, the LHS of \eqref{DELH2} also vanish. Now, we consider the spacetime to be stationary and axisymmetric. Therefore, the Killing vector $\chi^a$ can be written as $\chi^a=\chi^a_{(t)}+\Omega_H\chi^a_{(\phi)}$, where $\chi^a_{(t)}$ and $\chi^a_{(\phi)}$ are the components of the Killing vector $\chi^a$ along time and azimuthal directions respectively. With these considerations, the above equation \eqref{DELH2} yields as the first law in the Jordan frame \textit{i.e.}
 \begin{align}
\d M=T\d S+\Omega_H \d J~, \label{1STLAW}
\end{align}
where the temperature has been identified as $T=\kappa/(2\pi)$ and the other thermodynamic parameters are identified as \cite{Bhattacharya:2018xlq}
\begin{eqnarray}
&& \d S=\frac{\pi}{\kappa}\delta\int_{\mathcal{H}} d\Sigma_{ab}J'^{ab}[\chi]~;
\no 
\\
&& \d M=\frac{1}{2}\int_{\partial c_{\infty}} [\d(d\Sigma_{ab}J'^{ab}[\chi])-2d\Sigma_{ab}\chi^{[a}\T'^{b]}(q, \d q)]\Big|_{\chi=\chi_{(t)}}~;
\no 
\\
&& \d J=-\frac{1}{2}\int_{\partial c_{\infty}} [\d(d\Sigma_{ab}J'^{ab}[\chi])-2d\Sigma_{ab}\chi^{[a}\T'^{b]}(q, \d q)]\Big|_{\chi=\chi_{(\phi)}}~.
\no 
\\
 \label{SMJ}
\end{eqnarray}
We end up the discussion on the first law in the Jordan frame with the following discussions. Earlier, we have provided several arguments on why $L'$ serves as more appropriate Lagrangian than $L$ in the Jordan frame. Therefore, while obtaining first law, we have considered the Lagrangian in the Jordan frame as $L'$. However, if we had considered the Lagrangian in the Jordan frame as $L$, the thermodynamic 1st law could still be obtained in the form of Eq. \eqref{1STLAW}. However, in that case, the thermodynamic parameters (such as $S$, $M$, $J$), which are defined in the Eq. \eqref{SMJ}, would have been obtained in terms of $J^{ab}[\chi])$ and $\T^a(q,\d q)$ instead of $J'^{ab}[\chi])$ and $\T'^a(q,\d q)$. However, in that case, the thermodynamic parameters will not be exactly equivalent in the two frames (we shall discuss this in a greater detail later).
\vskip 1mm    
\noindent
{\bf Einstein frame:} In the Einstein frame, one can follow the same procedure \textit{i.e.} one can take the variation (due to the change of the field variables \textit{i.e.} $\t g_{ab}$ and $\t\phi$) of the conserved Noether current on-shell on the Killing horizon. Then, following the same algebraic steps as of the Jordan frame, one can obtain the first law in the Einstein frame as
\begin{align}
\d \t{M}=\t{T}\d \t{S}+\t{\Omega}_H \d \t{J}~, \label{1STLAWEIN}
\end{align}
where, the thermodynamic parameters in this frame will be defined as
\begin{eqnarray}
&& \d \t{S}=\frac{\pi}{\t{\kappa}}\delta\int_{\mathcal{H}} d\t{\Sigma}_{ab}\t{J}^{ab} [\t\chi]~;
\no 
\\
&& \d \t{M}=\frac{1}{2}\int_{\partial c_{\infty}}[\d( d\t{\Sigma}_{ab}\t{J}^{ab}[\t\chi])-2d\t\Sigma_{ab}\t{\chi}^{[a}\t{\T}^{b]}(\t{q}, \d \t{q})]\Big|_{\t{\chi}=\t{\chi}_{(t)}}~;
\no 
\\
&& \d \t{J}=-\frac{1}{2}\int_{\partial c_{\infty}} [\d(d\t{\Sigma}_{ab}\t{J}^{ab}[\t\chi])-2d\t\Sigma_{ab}\t{\chi}^{[a}\t{\T}^{b]}(\t{q}, \d \t{q})]\Big|_{\t{\chi}=\t{\chi}_{(\phi)}}~.
\no 
\\
 \label{SMJTIL}
\end{eqnarray}
Thus, we have obtained the first law in the two frames. In addition, we also have provided the covariant definition of the thermodynamic parameters in the two frames. In the following, we show how the thermodynamic parameters are related in the two frames.
\vskip 1mm    
\noindent
{\bf Comparison of the thermodynamic quantities in the two frames:} Earlier we have discussed that the imposition of the constraint \eqref{constraint} (or equivalently \eqref{constrainte}), which ensure that the Killing horizon in one frame is also a Killing horizon on the other, essentially means that the contravariant components of the Killing vector are the same in the two frames (as given by Eq. \eqref{CHICHITIL}).  As $\t{\chi^a}=\chi^a$, we obtain $\t{\chi_a}=\phi\chi_a$ and, the relation between the complementary null vectors in the two frames are given as $k_a=\t{k}_a$. Thus, 
\begin{align}
d\t{\Sigma}_{ab}=\sqrt{\t{\sigma}}(\t{\chi}_a\t{k}_b-\t{\chi}_b\t{k}_a)d^2x=\phi^2d\Sigma_{ab}~,\label{SIGMAREL}
\end{align}
where $\sigma$ and $\t{\sigma}=\phi^2\sigma$ denotes the determinant of the induced metric of the two-surfaces in the two frames (Jordan and Einstein) respectively. Furthermore, one can prove
\begin{align}
\t{J}^{ab}[\t\chi]=\frac{J'^{ab}[\chi]}{\phi^2}~. \label{JABJAB'}
\end{align}
and,
\begin{align}
\t{\Theta^a} (\t q, \d\t q)=\frac{\Theta'^a(q, \d q)}{\phi^2}~. \label{THTH'}
\end{align}
Using the above relations in \eqref{SMJ} and \eqref{SMJTIL}, one can straightforwardly obtain that the thermodynamic parameters (entropy, internal energy and angular momentum) are equivalent in the two frames. The equivalence of surface gravity has been discussed earlier while obtaining the zeroth law. Proof of the equivalence of the angular velocity is straightforward, which can be followed from \cite{Koga:1998un}. 

As we have mentioned earlier, if we had considered the Lagrangian in the Jordan frame as $L$ instead of $L'$ (as it has been done in the work of Koga and Maeda \cite{Koga:1998un}), we had obtained the thermodynamic parameters ($S$, $M$ and $J$) in terms of $J^{ab}[\chi]$ and $\Theta^a(q, \d q)$, in that case, the thermodynamic parameters ($S$, $M$ and $J$) cannot be shown to be exactly equivalent. In that case, the equivalence of these thermodynamic parameters are subject to some additional assumptions, such as the asymptotic flatness of the spacetime \cite{Koga:1998un}. However, in our approach no such assumptions are required. Therefore, it can be said that the $\square\phi$ term  which restores the relation \eqref{HOLNEW} in the Jordan frame, plays a crucial role in establishing the equivalence of the thermodynamic parameters in the two frames. Thus, the significance of the $\square\phi$ term has been missed earlier in the literature.

Finally, let us note that here we have obtained the thermodynamic laws using the Wald's formalism, and have defined the thermodynamic parameters in therms of the Noether current. On the other hand, in the ADT approach one can also obtain the thermodynamic law in a covariant manner, where one defines the thermodynamic parameters in terms of the ADT current. Earlier, we have shown that the Noether and the ADT currents are related by Eqs. \eqref{NOADT} and \eqref{NOADT'}. Although the ADT and Wald's formalisms (to obtain the first law in a covariant manner) have been developed independently, the relations \eqref{NOADT} and \eqref{NOADT'} show these two approaches are essentially the same. The proof of the first law in the ADT approach is established with the argument that the on-shell variation of the ADT current vanishes. This statement is equivalent to the Eq. \eqref{DELH1}, which has been obtained using Wald's approach. 

Note that for Killing horizon, the second law cannot be obtained as the black hole entropy, which is given by the area of the Killing horizon (or the $\phi A$ in the Jordan frame) does not change. 
In the following, we will discuss another approach to obtain thermodynamic interpretation of ST gravity.  Particularly one can find a thermodynamic structure of equation of motion for metric tensor on a generic null surface, where we shall discuss about the second law.

\subsection{Thermodynamic structure on a generic null surface}
In the earlier section, we have obtained the thermodynamic laws for the Killing horizon. In this section, we show that, like in GR, the gravitational equation of ST theory has a thermodynamic structure on a generic null surface. In conventional thermodynamics, the constancy of thermodynamic temperature (\textit{i.e.} the zeroth law) is obtained for a system in thermodynamic equilibrium. In black hole thermodynamics it corresponds to the system of Killing horizon. Therefore, for a generic null surface, we abstain from proving the zeroth law as we cannot argue how the idea of thermal equilibrium is justified for such system (\textit{i.e.} for the generic null hypersurface). Therefore, here we first describe the geometry of a null surface and then we move on to discuss about the first law and the second law in the two frames.

\subsubsection{Null geometry and the (1+3) foliation of a null surface} We consider that the whole $(1+3)$ dimensional spacetime manifold is $(\mathcal{M}, g_{ab})$. The null surface (denoted by $\mathcal{H}^{(N)}$), which is a three-dimensional hypersurface $(\mathcal{H}^{(N)}, \gamma_{\alpha\beta})$, is a submanifold lying inside $(\mathcal{M}, g_{ab})$. Now, the null surface is characterized by the fact that its tangent space is degenerate \textit{i.e.} if a vector $v^{\alpha}$ lies on the tangent plane of  $\mathcal{H}^{(N)}$, one obtains $\gamma_{\alpha\beta}v^{\alpha}=0$. For this reason (that the tangent space is degenerate), it is impossible to define a projection operator, which projects every vector onto its tangent plane. In addition, the null surface is generated by the null geodesic congruences. Therefore, we denote the normal to $\mathcal{H}^{(N)}$ as $l^a$ (that generates the null surface), which is a null vector and obeys the geodesic equation $l^a \na_al^b=\kappa l^b$ \cite{Gourgoulhon:2005ng}. Here $\kappa$ is the non-affinity parameter. For black hole horizon, it can be identified as the surface gravity of the black hole. Since the null surface $\mathcal{H}^{(N)}$ is self-orthogonal, it will be useful to introduce the auxiliary null vector $k^a$, which is defined by the condition $l^ak_a=-1$.

Although we cannot define a projection tensor (or induced metric) for $\mathcal{H}^{(N)}$, we can foliate $\mathcal{H}^{(N)}$ in terms of a family of two-dimensional spacelike hypersurfaces and can define a projection vector onto it. We consider the whole manifold $(\mathcal{M}, g_{ab})$ is foliated in terms of $t=$const. hypersurfaces, which are denoted by $\Sigma_t$. The intersection of $\Sigma_t$ and $\mathcal{H}^{(N)}$ will be a two dimensional hypersurface $\mathcal{S}_t$ (i.e. $\mathcal{S}_t=\mathcal{H}^{(N)}\cap \Sigma_t$). The induced metric on $\mathcal{S}_t$ can be given in terms of $l^a$ and $k^a$ as
\begin{align}
q_{ab}=g_{ab}+l_ak_b+l_bk_a~. \label{QAB2}
\end{align}

As we have discussed earlier, the conformal transformation maps a null surface $\mathcal{H}^{(N)}$ (Jordan frame) to a null surface $\t{\mathcal{H}}^{(N)}$ (Einstein frame). The null surface of the conformal frame (or the Einsten frame) will be will be generated by the null vector $\t l^a$, which obeys the geodesic equation $\t l^a \t\na_a\t l^b=\t \kappa \t l^b$. Furthermore, the induced metric on the spatial cross-section of $\t{\mathcal{H}}^{(N)}$ (\textit{i.e.} on $\t {\mathcal{S}}_t$) will be given as
\begin{align}
\t q_{ab}=\t g_{ab}+\t l_a\t k_b+\t l_b\t k_a~. \label{QAB2TIL}
\end{align}
The components of null vectors are connected in the two frames as \cite{Bhattacharya:2020wdl}
\begin{align}
\t l^a&=l^a, \ \ \ \ \ \ \ \ \ \  \t l_a=\phi l_a &
\no 
\\
\t k^a&=\frac{1}{\phi} k^a, \ \ \ \ \ \  \t k_a= k_a~.& \label{LNK}
\end{align}
Now we move on to obtain the first law of a generic null hypersurface in the two frames

\subsubsection{Thermodynamic structure of ST gravity} \label{1STLAWGNC}
It has been found earlier (in the context of GR \cite{Chakraborty:2015aja, Padmanabhan:2002sha, Kothawala:2007em} and Lanczos-Lovelock gravity \cite{Paranjape:2006ca, Kothawala:2009kc, Chakraborty:2015wma}) that the dynamic equation projected on the horizon provides the equilibrium version provides the expression of first law. In that case, the change in thermodynamic parameters are arises due to the virtual change of the of the horizon radius (say from $r_H$ to $r_H+\d r_H$). Later, for a generic null surface, the same law has been obtained by taking a proper projection of the dynamical equation on the horizon surface \cite{Chakraborty:2015aja,Chakraborty:2015wma, Kothawala:2010bf,Dey:2020tkj}. In that case, the change in thermodynamic parameters arise due to the virtual displacement of the horizon along the affine parameter ($\lambda_{(k)}$). In this case, the dynamical equation in the Jordan frame is given as $E_{ab}=T_{ab}^{(m)}/2$ (in the Einstein frame, it is $\t E_{ab}=\t T_{ab}^{(m)}/2$)~. It can be shown, the projection of $E_{ab}$ on $\mathcal{H}^{(N)}$ provides several interesting implications.
\begin{itemize}
\item $E_{ab}l^al^b$ (\textit{i.e.} the projection of $E_{ab}$ along the normal of the $\mathcal{H}^{(N)}$) is related to the second law of black hole thermodynamics. Later, when we prove the generalized second law, it will be shown that this projection will play the crucial role to prove the entropy increase theorem.
\item $E_{ab}l^aq^b_c$ is related to the fluid-gravity correspondence in scalar tensor gravity \cite{Bhattacharya:2020wdl}.
\item $E_{ab}l^ak^b$ is related to the thermodynamic description of a generic null surface in scalar tensor gravity. Thus, in the present case, this projection will be used to obtain the first law.
\end{itemize}
Although we have discussed about projection of $E_{ab}$ only, the same comments are valid for $\t E_{ab}$ as well. In the following, we discuss the procedure to obtain the first law like structure of a generic null hypersurface using the projections $E_{ab}l^ak^b$ and $\t E_{ab}\t l^a\t k^b$ in the two frames.

The finding of the 1st law like structure in the two frames largely depends upon the following geometrical identity (see \cite{Dey:2020tkj}, where this identity has been obtained).
\begin{align}
-\kappa\theta_{({k})}=- D_a\O^a-\O_a\O^a+\theta_{({l})}\theta_{({k})}+l^i\na_i\theta_{({k})}+\frac{1}{2}\  ^{(2)}R-R_{ab} l^a k^b-\frac{1}{2} R~,
\label{RELTNJOR}
\end{align}
where ${{D_a}}$ is the covariant derivative operator defined on the manifold $({S}_t, {q}_{ab})$ and $^{(2)} {R}$
denotes the Ricci scalar associated with the operator ${{D_a}}$. The proof of the 1st law in the Einstein frame is comparatively simpler than the Jordan frame. Therefore, for simplicity, first we start the analysis in the Einstein frame. 
\vskip 1mm    
\noindent
{\bf Einstein frame:} In the Einstein frame, the expression of the above identity \eqref{RELTNJOR} is essentially the same, albeit defined in terms of the tilde variables
\begin{align}
-\t\kappa\t\theta_{(\tilde{k})}=-\t D_a\t\O^a-\t\O_a\t\O^a+\t\theta_{(\tilde{l})}\t\theta_{(\tilde{k})}+\t l^i\t\na_i\t\theta_{(\tilde{k})}+\frac{1}{2}\  ^{(2)}\t R-\t R_{ab}\t l^a\t k^b-\frac{1}{2}\t R~.
\label{RELTN}
\end{align}
One can replace the last two terms of \eqref{RELTN} using the projection $\t E_{ab}\t l^a\t k^b$ and can obtain the 1st law in the following manner. We consider a virtual displacement of the null surface along the direction of the auxiliary null vector $\t k^a$. Firstly, let us consider $\t k^a$ is parametrized by $\lambda_{(\t k)}$, which implies $\t k^i=-dx^i/d\lambda_{(\t k)}$. Here, we put the extra negative sign in the definition of $\t k^i$ for the following reasons. Remember, that the auxiliary null vector $\t k^i$ also corresponds to the ingoing null vector (\textit{i.e.} $x^i$ decreases for the increase of $\lambda_{(\t k)}$). Therefore, in order to make the change of the thermodynamic quantities positive along the virtual displacement along $\t k^i$, we defined $\t k^i$ with the negative sign. The virtual displacement along $\t k^i$ can be explained in the following manner. We consider two null surfaces are located at $\lambda_{(\t k)}=0$ and at $\lambda_{(\t k)}=\d \lambda_{(\t k)}$. A virtual displacement along $\t k^i$ implies a shift from one solution of null hypersurface (located at $\lambda_{(\t k)}=0$) to the other (located at $\lambda_{(\t k)}=\d \lambda_{(\t k)}$). Under this virtual displacement, the above identity \eqref{RELTN} can be identified as the 1st law of thermodynamics for the generic null surface. For that, we multiply both sides of the Eq. \eqref{RELTN} with $\d \lambda_{(\t k)}$ (accounting an overall factor of $1/8\pi$) and integrate it over the two-surface $\tilde{\mathcal{S}}_{t}$. Also, we use the projection $\t E_{ab}\t l^a\t k^b=\t T_{ab}^{(m)}\t l^a\t k^b/2$ and finally obtain
\begin{align}
-\int_{\tilde{S}_t} d^2 x \sqrt{\t q}  ~ \delta \lambda_{(\t k)} \frac{\t \kappa}{2 \pi} \frac{1}{4}\t\theta_{(\tilde{k})}=  \int_{\tilde{S}_t} d^2 x \sqrt{\t q}~\delta \lambda_{(\t k)} \frac{1}{8 \pi}\Big[\frac{1}{2}{^2 \t R} +\t l^i \t  \nabla_i \t\theta_{(\tilde{k})}+\t\theta_{(\tilde{l})}\t\theta_{(\tilde{k})} -\t \Omega_a \t \Omega^a - \t D_A\t \Omega^A\Big] 
\nonumber
\\
- \int_{\tilde{S}_t} d^2 x \sqrt{\t q}~\delta \lambda_{(\t k)}\Big[\t T_{ab}^{(\t\phi)}+\t T_{ab}^{(m)} \Big]\t l^a \t k^b  ~. \label{RELTHER}
\end{align}
Here, $\t T_{ab}^{(\t\phi)}$ corresponds to the Energy-momentum tensor of the scalar field $\phi$, which is given as
 \begin{align}
 \t T_{ab}^{(\t \phi)}=\tilde{\nabla}_a\tilde{\phi}\tilde{\nabla}_b\tilde{\phi}-\frac{1}{2}\tilde{g}_{ab}\tilde{\nabla}^i\tilde{\phi}\tilde{\nabla}_i\tilde{\phi}-\tilde{g}_{ab}U(\tilde{\phi})~. \label{TABPHITIL}
 \end{align}
The above expression of Eq. \eqref{RELTHER} can be expressed as the 1st law on the null surface of the following form
\begin{align}
\int_{\tilde{S}_t} d^2 x\t T \delta_{\lambda(\t k)} \t s = \delta_{\lambda(\t k)} \t E + \t F \delta \lambda_{(\t k)} ~,\label{1STLAWEINNULL}
\end{align}
Here, the thermodynamic parameters are identified as follows. The temperature $\t T$ is identified as $\t T=\t\kappa/2\pi$, the entropy density $\t s$ is identified as $\t s=\sqrt{\t q}/4$ (thus, the total entropy is quarter of the horizon area \textit{i.e.} $\t S=\int_{\tilde{S}_t} d^2 x \t s=\frac{1}{4}\int_{\tilde{S}_t} \sqrt{\t q} d^2 x$). Here $\delta_{\lambda(\t k)} \t s$ corresponds to the change of entropy density due to the virtual displacement along $\t k$, which is explicitly given as
\begin{align}
\delta_{\lambda(\t k)} \t s =\frac{d\t s}{d \lambda_{(\t k)}} \delta\lambda_{(\t k)} =\frac{1}{4}\frac{d\sqrt{\t q}}{d \lambda_{(\t k)}} \delta\lambda_{(\t k)}=-\frac{1}{4}\sqrt{\t q} ~\t\theta_{(\t{k})}\delta\lambda_{(\t k)}~.\label{CHNGENTROPYDEN}
\end{align}
Similarly, $\delta_{\lambda(\t k)} \t E$ (in \eqref{1STLAWEINNULL}) implies the change of energy due to the mentioned virtual displacement, which is given as
\begin{align}
\delta_{\lambda(\t k)} \t E =\frac{1}{8 \pi} \int_{\tilde{S}_t} d^2 x \sqrt{\t q} ~ \delta \lambda_{(\t k)}\Big[\frac{1}{2}{^2 \t R} +\t l^i \t  \nabla_i \t\theta_{(\tilde{k})}+\t\theta_{(\tilde{l})}\t\theta_{(\tilde{k})} -\t \Omega_a \t \Omega^a - \t D_A\t \Omega^A\Big] ~.
\label{varEe}
\end{align}
The expression of energy $\t E$ (associated with the two surface $S_t$) can be obtained performing an indefinite integration over the affine length $\delta \lambda_{(\t k)}$, which is given as
\begin{align}
\t E =\frac{1}{8 \pi} \int_{\tilde{S}_t} \int d^2 x \sqrt{\t q} ~d \lambda_{(\t k)}\Big[\frac{1}{2}{^2 \t R} +\t l^i \t  \nabla_i \t\theta_{(\tilde{k})}+\t\theta_{(\tilde{l})}\t\theta_{(\tilde{k})} -\t \Omega_a \t \Omega^a - \t D_A\t \Omega^A\Big] ~.
\label{Ee}
\end{align}
Note that the above expression of energy resembles significantly to the expression of Hawking-Heyward quasi-local energy \cite{Hayward:1997jp, Prain:2015tda}. In addition, the identification of the above expression \eqref{Ee} as energy because it indeed provides the expression of energy for the well-known spacetimes \cite{Dey:2020tkj,Chakraborty:2015aja}.

Finally, we identify the thermodynamic pressure ($\t P$) as $\t P=-(\t T_{ab}^{(\t\phi)}+\t T_{ab}^{(m)})\t l^a \t k^b$, which is inspired from the earlier work \cite{Hayward:1997jp} and recently been defined in \cite{Chakraborty:2015aja,Chakraborty:2015wma, Kothawala:2010bf} in a similar way. The work term ($\t W$) due to the virtual displacement $\delta\lambda_{(\t k)}$ is, thus, obtained as  
\begin{align}
\t W=\t F\delta\lambda_{(\t k)}=\int_{\tilde{S}_t} d^2x \sqrt{\t q} ~\delta\lambda_{(\t k)} \t P =-\int_{\tilde{S}_t} d^2x \sqrt{\t q}~\delta\lambda_{(\t k)} (\t T_{ab}^{(\t\phi)}+\t T_{ab}^{(m)})\t l^a \t k^b ~,
\end{align}
where $\t F$, which is the pressure integrated over the two-surface $S_t$, can be interpreted as the generalized force conjugate to the virtual displacement $\delta\lambda_{(\t k)}$. 

Thus, in this section we show that the relevant projection of $\t E_{ab}$ on to the null surface (\textit{i.e.} $\t E_{ab}\t l^a\t k^b$), give rise to the expression of the first law. In the following, we aim to establish the similar thermodynamic interpretation in the Jordan frame. Before that, let us note that here the thermodynamic parameters are are defined in terms of several parameters like $\t\kappa$ $\t\theta_{(\tilde{l})}$, $\t\theta_{(\tilde{k})}$, $\t \Omega_a$ \textit{etc.}, which are related to the parameters of the Jordan frame ($\kappa$, $\theta_{(l)}$, $\theta_{(k)}$, $\Omega_a$ \textit{etc.}) as \cite{Bhattacharya:2020wdl,Dey:2021rke}
\begin{align}
\t\th_{(\t{l})}&=\th_{({l})}+l^i\na_i\ln\phi~,
\no 
\\
\t\th_{(\t{k})}&=\frac{1}{\phi}\Big[\th_{({k})}+k^i\na_i\ln\phi\Big],
\no 
\\
\t\k&=\k+l^i\na_i\ln\phi~,
\no 
\\
\t\o_a&=\o_a+\frac{1}{2}\Big[l_ak^i\na_i\ln\phi+\na_a\ln\phi-k_al^i\na_i\ln\phi\Big]~,
\no 
\\
\t\O_a&=\O_a+\frac{1}{2}q^b_a\na_b\ln\phi~.
\label{TRANSREL}
\end{align}
With these, we now move forward to obtain the first law of a generic null surface in the Jordan frame.
\vskip 1mm    
\noindent
{\bf Jordan frame:} As we have mentioned earlier, the major challenge which arises in the  formulation of thermodynamic law in the scalar-tensor gravity, is taking the stand on whether the thermodynamic parameters in the two frames are conformally invariant. Earlier, for the Killing horizon, we have established that the thermodynamic parameters are exactly equivalent in the two frames. Therefore, we expect the same to hold for the generic null surface as well. The thermodynamic 1st law and the thermodynamic parameters are consistently obtained for a generic null surface in the Einstein frame. Therefore, here our goal is obtaining the first law in such a way that the defined thermodynamic parameters becomes conformally equivalent to that of the Einstein frame. This is done in the following manner. Inspired by the method which we earlier developed to obtain fluid-gravity correspondence \cite{Bhattacharya:2020wdl}, we defined the thermodynamic parameters of the Jordan frame in terms of the parameters of the Einstein frame (such as $\t\theta_{(\tilde{l})}$, $\t\theta_{(\tilde{k})}$, $\t \Omega_a$ \textit{etc.}) in the background of the Jordan frame ($g_{ab}$, $q^a_b$, $\nabla_a$, $D_A$ \textit{etc.}). The relevant relation in the Jordan frame can be obtained as \cite{Dey:2021rke}
\begin{eqnarray}
&&-\t\kappa\t\theta_{(\t{k})}=- D_a\t\O^a-\t\O^i\na_i(\ln\phi)-\t\O_a\t\O^a+\t\theta_{(\t{l})}\t\theta_{(\t{k})}+ l^i\na_i\t\theta_{(\t{k})}+\frac{1}{2\phi}\  ^{(2)} R-\frac{1}{2\phi}D^iD_i(\ln\phi)
\no 
\\
&&-\Big(\frac{R_{ab} l^a k^b}{\phi}+\frac{R}{2\phi}+\frac{3}{2\phi}l^ak^b\na_a(\ln\phi)\na_b(\ln\phi)-\frac{l^ak^b}{\phi^2}\na_a\na_b\phi-\frac{1}{\phi^2}\nabla^i\na_i\phi +\frac{3}{4\phi}\na_i(\ln\phi)\na^i(\ln\phi)\Big)~.
\no 
\\ \label{RELATHER}
\end{eqnarray}
Now, using the projection of $E_{ab}$ (\textit{i.e.} $E_{ab}l^ak^b=T_{ab}^{(m)}l^ak^b/2$) one obtains
\begin{eqnarray}
&&-\phi \t\kappa\t\theta_{(\t{k})}=-\phi D_a\t\O^a-\t\O^i\na_i(\ln\phi)- \phi \t\O_a\t\O^a+ \phi \t\theta_{(\t{l})}\t\theta_{(\t{k})}+ \phi l^i\na_i\t\theta_{(\t{k})}+\frac{1}{2}\  ^{(2)} R-\frac{1}{2}D^iD_i(\ln\phi)
\no 
\\
&&-{l^ak^b}\Big[\Big(\frac{2\omega+3}{2}\Big)\Big\{\na_a(\ln\phi)\na_b(\ln\phi)-\frac{1}{2}g_{ab}\na^i(\ln\phi)\na_i(\ln\phi)\Big\}-\frac{V}{2\phi}g_{ab}\Big] -\frac{8\pi}{\phi}T_{ab}^{(m)}l^ak^b~. \label{RELWITHT}
\end{eqnarray}
Now, we have obtained earlier the expression of $\t T_{ab}^{(\t \phi)}$ in Eq. \eqref{TABPHITIL}, which under the transformation relations \eqref{GAB} and \eqref{PHI} can be written equivalently as
\begin{align}
\t T_{ab}^{(\t \phi)}\equiv \Big(\frac{2\omega+3}{16\pi}\Big)\Big\{\na_a(\ln\phi)\na_b(\ln\phi)-\frac{1}{2}g_{ab}\na^i(\ln\phi)\na_i(\ln\phi)\Big\}-\frac{V}{16\pi\phi}g_{ab} \label{TABPHIEQUIV}
\end{align}
In addition, the virtual displacement by a small amount of affine length is related to the virtual displacement in the spatial coordinates as
\begin{equation}
\delta{x^a} = -\t{k^a} \delta{\lambda_{(\t k)}}  = -\frac{k^a}{\phi} \delta{\lambda_{(\t k)}} = -k^a \delta{\lambda_{k}}~.   
\end{equation}
This implies that the virtual displacement by an amount  $\delta{\lambda_{(\t k)}}$ is equivalent to the displacement $\phi ~\delta{\lambda_{k}}$ in the Einstein frame. Therefore, we multiply the both sides of Eq. \eqref{RELWITHT} by $\phi ~\delta{\lambda_{k}}$ and integrate over the two-surface $\mathcal{S}_t$, which yields
\begin{eqnarray}
&&-\int_{S_t} d^2 x\sqrt{q} ~\delta \lambda_{(k)}\phi^2  \frac{\t \kappa}{2 \pi} \frac{1}{4}\t\theta_{(\t{k})} = -\int_{S_t} d^2 x \sqrt{q}~\delta \lambda_{(k)} \phi\Big[\t T_{ab}^{(\t\phi)}+\frac{ T_{ab}^{(m)}}{\phi} \Big] l^a k^b 
\nonumber
\\
&&+\int_{S_t} d^2 x \sqrt{q}~\delta \lambda_{(k)} \frac{\phi^2}{8 \pi}\Big[\frac{1}{2\phi}{^{(2)}  R} +l^i \nabla_i \t\theta_{(\t{k})}+\t\theta_{(\t{l})}\t\theta_{(\t{k})} -\t \Omega_a \t \Omega^a - \t D_A\t \Omega^A-\t\O^i\na_i(\ln\phi)
\nonumber 
\\
&&-\frac{1}{2\phi}D^iD_i(\ln\phi)\Big]  ~.
\label{RELTHERJOR} 
\end{eqnarray}
The above relation \eqref{RELTHERJOR} can be interpreted as the 1st law in the Jordan frame, which is given as
\begin{align}
\int_{S_t} d^2 x T \delta_{\lambda(k)} s = \delta_{\lambda(k)} E +  F \delta \lambda_{(k)} ~.\label{1STLAWJOR}
\end{align}
In this frame, the thermodynamic parameters are identified as follows. The temperature in this frame is identified as $T=\kappa/2\pi$~, which have been proven to be equivalent to $\t T$ for the Killing horizon. The entropy density is identified as $s=\phi\sqrt{q}/4$, so that the total entropy is defined as 
\begin{align}
S=\int_{S_t}sd^2x=\int_{S_t}\phi \frac{\sqrt{q}}{4} d^2x= \int_{\t{S_t}} \frac{\sqrt{\t{q}}}{4} d^2 x = \t S~,  \label{ENTJOR}
\end{align}
as $\sqrt{\t q}=\phi\sqrt{q}$. The change of entropy density due to the virtual displacement of the null surface is defined as $\delta_{\lambda(k)} s$, which can be explicitly obtained as 
\begin{align}
\delta_{\lambda(k)} s =\frac{ds}{d\lambda_{(k)}}\delta\lambda_{(k)}=\frac{\delta\lambda_{(k)}}{4}\Big(\phi\frac{d\sqrt{q}}{d\lambda_{(k)}}+\sqrt{q}\frac{d\phi}{d\lambda_{(k)}}\Big)=-\delta\lambda_{(k)}\frac{\sqrt{q}\phi}{4}\Big(\theta_{(k)}+k^i\na_i(\ln\phi)\Big)
\nonumber
\\
=-\frac{1}{4}\phi^2\sqrt{q}\t\theta_{(\t{k})}\delta\lambda_{(k)} = -\frac{1}{4}\sqrt{\t{q}} ~\t{\theta}_{(\t{k})} \delta{\lambda_{(\t{k})}} = \delta_{\lambda(\t{k})} \t{s}~, \ \ \ \ \ \ \ \ \ \ \ \ \ \ \ 
\end{align}
The change in energy $E$ (due to the virtual displacement) can be obtained as
\begin{align}
\d_{\lambda{(k)}}E=\frac{1}{8 \pi}  \int_{S_t} d^2 x \sqrt{q} ~\delta\lambda_{(k)} \phi^2\Big[\frac{1}{2\phi}{^{(2)}  R} +l^i \nabla_i \t\theta_{(\t{k})}+\t\theta_{(\t{l})}\t\theta_{(\t{k})} -\t \Omega_a \t \Omega^a - \t D_A\t \Omega^A
\nonumber
\\
-\t\O^i\na_i(\ln\phi)-\frac{1}{2\phi}D^iD_i(\ln\phi)\Big] ~,
\label{varEj}
\end{align} 
and the total energy associated to $S_t$ can be obtained as
\begin{align}
E=\frac{1}{8 \pi} \int_{S_t} \int d^2 x \sqrt{q} ~d\lambda_{(k)} \phi^2\Big[\frac{1}{2\phi}{^{(2)}  R} +l^i \nabla_i \t\theta_{(\t{k})}+\t\theta_{(\t{l})}\t\theta_{(\t{k})} -\t \Omega_a \t \Omega^a - \t D_A\t \Omega^A
\nonumber
\\
-\t\O^i\na_i(\ln\phi)-\frac{1}{2\phi}D^iD_i(\ln\phi)\Big]  ~.
\label{Ej}
\end{align}
Thus, due to the presence of the scalar field $\phi$ (which is non-minimally coupled), the expression of energy and its variation has a bit different expression as compared to that of the Einstein frame. However, using the conformal transformation relations, one can obtain that both the energy and its variation is equivalent in the two frames (\textit{i.e.} $E=\t E$ and $\d_{\lambda{(k)}}E=\delta_{\lambda(\t k)} \t E$).

The work term is defined in the following manner. Firstly, the pressure ($P$) is identified as $P = - (\phi \t{T}^{(\t{\phi})}_{ab} + T_{ab}^{(m)}) l^a k^b$ (thus, $ P=\phi^2 \t P$). The expression of work done (under the virtual displacement) is obtained in the same way as of the Einstein frame, which is given as 
\begin{align}
W= F \delta \lambda_{(k)}=-\int_{S_t} d^2 x ~\sqrt{q} ~\delta\lambda_{(k)} \Big({\phi\t T_{ab}^{(\t\phi)}}+T_{ab}^{(m)}\Big) l^a k^b~,
\end{align}
where the generalized force term $F$ (which is conjugate to the virtual displacement $\delta \lambda_{(k)}$) is identified as $F = \int_{S_t} d^2 x \sqrt{q} P$ (which implies $F=\phi\t F$). Thus, one can find that the work done due to the virtual displacement are the equivalent \textit{i.e.} $W=\t W$.

Thus, in this section, we have obtained the first law in the two frames and have obtained the fact that the thermodynamic parameters are equivalent in the two frames, which have been a major challenge for long. In the following, we obtain the entropy increase theorem (\textit{i.e.} the second law) in the two frames.

There are two well known ways of obtaining the first law for black holes: (i) The physical process version, in which the black hole parameters actually change by a physical process (such as the increase of the black hole mass by absorbing more mass) (ii) the equilibrium version/ stationary state version, in which black hole parameters do not change but we just compare two (stationary) black holes with parameters being infinitesimally close to each other. The method which we have adopted (originally developed by Padmanabhan et. al. \cite{Chakraborty:2015aja}) evolves around the concept of virtual displacement and it resembles to the equilibrium version. Here, we compare two null surfaces, locations of which are separated by $\delta x^i=-k^i\delta\lambda_{(k)}$. Due to this virtual displacement in the location, we observed that the equation of motion projected on the null surface has a thermodynamic structure which has resemblance with the first law of thermodynamics. By comparing with the usual law of thermodynamics we identified the change in thermodynamic quantities.

Another important comment is as follows. As we know, there are no unique way of identifying energy in general relativity. As a result, several prescriptions are provided in the literature such as ADM energy, Misner-Sharp energy, Hawking-Hayward energy, Brown-York energy \textit{etc.} Our definition of energy (in Eq. (\ref{Ee})) looks very similar to the Hawking-Hayward energy. However, the Hawking-Hayward energy is not conformally invariant (and therefore cannot be the probable candidate of energy in our case for the reasons described in the manuscript), whereas, the energy which we have defined, is conformally invariant. Furthermore, for the metric in Gaussian null coordinates (GNC), our definition of energy boils down to the expression of energy defined in the work of T. Padmanabhan \cite{Chakraborty:2015aja}.

\subsubsection{The second law} Earlier, we have obtained the expression of entropy in the two frames. Here, it will be proved that the change in entropy (either by any physical process or any near-equilibrium change) is always positive quantity, \textit{i.e.} entropy always increases. The proof of the entropy increase theorem crucially depends on the following three aspects: (i) The null Raychaudhuri equation (NRE), (ii) The projection $E_{ab} l^al^b$ (or $\t E_{ab} \t l^a\t l^b$)  and (iii) Null energy condition \textit{i.e.} $T^{(m)}_{~ab}l^al^b\geq 0$ (or equivalently $\t T^{(m)}_{~ab}\t l^a\t l^b\geq 0$; as $\t T^{(m)}_{~ab}$ and $T^{(m)}_{~ab}$ are related by \eqref{conformtab}, where as $\t l^a$ and $l^a$ are related by \eqref{LNK}). In the following, we establish the entropy increase theorem in the two frames. For simplicity, we first do our analysis in the Einstein frame, then the same analysis is performed for the Jordan frame.
\vskip 1mm    
\noindent
{\bf Einstein frame:} The null horizon is generated by the null geodesic congruences, where the generator of the null surface $\t l^a$ satisfies the geodesic condition $\t l^a\t\na_a \t l^b=\t\k \t k^b$. Thus, the change in entropy along the null generator ($\t l^a$) is measured by 
\begin{align}
\frac{d\t S}{d\llt}=\frac{1}{4}\int_{\t{\mathcal{S}}_t} \frac{\sqrt{\t q}}{d\llt} d^2 x=\frac{1}{4}\int_{\tilde{\mathcal{S}}_t} \sqrt{\t q} ~\t\th_{{(\t l)}}d^2 x \label{entein1}
\end{align}
Thus, from \eqref{entein1}, it can be concluded that the entropy can decrease only when $\t\th_{{(\t l)}}$ is negative.
Now, the null Raychaudhuri equation (NRE ) is given as
\begin{align}
\frac{d \t\th_{{(\t l)}}}{d \llt}=\kt\thlt-\t\s_{ab}\t\s^{ab}-\frac{1}{2}\thlt^2-\t R_{ab}\t l^a\t l^b~. \label{RAYGR2}
\end{align}
Using the projection of $\t E_{ab}\t l^a\t l^b$ in \eqref{RAYGR2}, one can obtain
\begin{align}
\frac{d \t\th_{{(\t l)}}}{d \llt}=\kt\thlt-\t\s_{ab}\t\s^{ab}-\frac{1}{2}\thlt^2-8\pi (\t T_{ab}^{\t\phi}+\t T_{ab}^{(m)})\t l^a\t l^b~. \label{RAYEIN}
\end{align}
One can find that $\t T_{ab}^{\t\phi}\t l^a\t l^b=(\t l^a\p_a\t\phi)^2$, which is a positive definite. Now, we consider that the external matter source obeys the null energy condition $\t T^{(m)}_{~ab}\t l^a\t l^b\geq 0$~. Therefore, all the terms on the RHS of \eqref{RAYEIN} is negative except the term $\kt\thlt$~. Now, if $\thlt$ is initially negative, then it implies $\frac{d \t\th_{{(\t l)}}}{d \llt}< 0$, which means $\thlt$ will further decrease and become $-\infty$ within finite $\llt$. Thus, ultimately it will lead to the formation of caustic. Hence, in order to assure that there is no caustic in future $\thlt$ will have to be positive always. This, in Eq. \eqref{entein1} implies that the entropy cannot decrease and we have
\begin{align}
\frac{d\t S}{d\llt}\geq 0~,
\end{align}
which is the second law in the Einstein frame.
\vskip 1mm    
\noindent
{\bf Jordan frame:} In the Jordan frame, the entropy increase theorem can be proven following the similar steps as of the Einstein frame. In this frame, the entropy is given as $S=\int_{S_t}\phi \frac{\sqrt{q}}{4} d^2x$. Thus, the change in entropy along $l^a$ is given as
\begin{align}
\frac{dS}{d\lambda}=\frac{1}{4}\int_{\mathcal{S}_t}\sqrt{q}\Big(\phi\theta+l^i\na_i\phi \Big)d^2x=\frac{1}{4}\int_{\mathcal{S}_t}\sqrt{q}\phi\t\th_{(\t l)} d^2x~. \label{CHNGENTROPYJOR}
\end{align}
Thus, the change of entropy in the Jordan frame is determined in terms of the parameter $\thlt$ and not $\thl$~. The fact that $\thlt$ cannot be negative, has been proven earlier in the Einstein frame. Nevertheless, the positivity of $\thlt$ can be proven independently using the null Raychaudhuri equation (NRE) of the Jordan frame, the projection $E_{ab}l^al^b$ and the null energy condition of the Jordan frame (\textit{i.e.} $T_{ab}^{(m)}l^al^b>0$) \cite{Bhattacharya:2018xlq,Bhattacharya:2020wdl}. This is possible because (i) NRE is a geometric identity and the similar expression holds in both the frames (either written in terms of tilde variables of the Einstein frame or in terms of non-tilde variables of the Jordan frame), (ii) The dynamical equations $E_{ab}$ and $\t E_{ab}$ are equivalent (\textit{i.e.} $E_{ab}=\phi\t E_{ab}$) and (iii) The null energy condition in one frame implies the same of another frame (as the energy momentum tensor in one frame in related to the same in the other frame by Eq. \eqref{conformtab} and the null vectors are related by \eqref{LNK}). In any case, one can finally obtain 
\begin{align}
\frac{dS}{d\lambda}\geq 0~,
\end{align}
which is the entropy increase theorem in the Jordan frame.

Finally, it is worth to point out that no caustic is one of the key argument to prove the second law. In this regard we mention that the argument of no caustic is valid for event horizon and for a generic null surface with the generators having no future end points (as in general caustics may appear for a null surface). Thus, the proof of the second law can be attributed to those null surfaces, the generators of which do not have any future end points (such as the event horizon of a black hole).
%%%%%%%%%%%%%%%%%%%%%%%%%%%%%%%%%%%%%%%%%%%%%%%%%%%%%%%%%%%%%%%%%%%%%%%%%%%%%%
%%%%%%%%%%%%%%%%%%%%%%%%%%%%%%%%%%%%%%%%%%%%%%%%%%%%%%%%%%%%%%%%%%%%%%%%%%%%%
\section{Fluid-gravity correspondence}
In the earlier section, we have extensively discussed about the thermodynamics in the two frames of the scalar-tensor gravity. In that case, we have explicitly mentioned how, in addition to the proper formulation of thermodynamics,  one has to worry about whether the thermodynamic parameters are conformally invariant. The analogy of gravity with fluid dynamics is quite an old finding. It was shown by Damour \cite{DAMOUR} that the Einstein's equation, when projected on the null surface $(G_{ab}l^aq^b_c$), provides an analogous equation of the non-relativistic Navier-Stokes equation. This work was in consistent with the idea of the black hole sear viscosity proposed by Hawking and Hartle \cite{Hartle:1973zz,Hartle:1974gy,Hawking:1972hy}. This finding by Damour later paved the idea of the membrane paradigm of black hole horizon. Here we briefly discuss that the fluid-gravity analogy can be established in scalar-tensor gravity as well following the footsteps of Damour. However, in this case, one can draw two pictures of fluid-gravity connection: (i) The equivalent picture, where the thermodynamic parameters are equivalent but, it may violate the well-known Kovtun-Son-Starinets (KSS) bound and (ii) The inequivalent picture where the KSS bound is maintained but the fluid parameters are not conformally invariant. In the following, we provide fluid-gravity correspondence in scalar-tensor gravity and discuss these two pictures in more detail.

Before going to the main discussion let us pointed out that there are two radically different approaches of fluid-gravity, one is via AdS/CFT and the other one via projection of the Einstein's equation on the null surface, which is inspired by the membrane paradigm. However, in literature, both the approaches are known as the ``fluid-gravity correspondence''. This is why we also call it ``fluid-gravity correspondence in the scalar-tensor gravity'', while our method follows from the second approach (projection of dynamical equation). In the AdS/CFT approach, the bulk metric geometry is constructed by using the fluid data living at the asymptotic boundary such that the Brown-York stress tensor of the geometry yields the required fluid equation on this boundary. Whereas the second approach, which we followed here, is completely different from the earlier one and implies different significance

Note that the approach, which we discuss in the following, has an enormous significance in the context of interpreting gravity as an emergent phenomenon (which has been advocated in the work of T. Padmanabhan as well, such as \cite{Padmanabhan:2010rp}). Earlier it was shown that the Einstein's equation, when it is projected on the null surface, takes the form of fluid-dynamic equation (\textit{i.e.} the Navier-Stokes equation), which implies as Einstein's equation has the same status as of the equations of fluid dynamics. Therefore, gravity can be interpreted as an emergent phenomenon like, say, fluid mechanics. In addition, although these two approaches are different, they converge to the same expression of the viscosity coefficients and the KSS bounds. Thus both of these approaches are the well-accepted formalisms in the literature. In our manuscript, we have extended the second approach for the scalar-tensor and $f(R)$ gravity.

 The fluid-gravity analogy in the present context is established from the following geometrical identity \cite{Padmanabhan:2010rp}
 \begin{align}
R_{mn} l^m q^n_a= q^n_a\liel\O_n+\th^{(l)}\O_a- D_a\Big(\frac{\th^{(l)}}{2}+\k\Big)+ D_i\sigma^i_a~, \label{NOTDNS1}
\end{align} 
In the following, we obtain the fluid-gravity correspondence in the two frames. In Einstein frame, the fluid-gravity correspondence can be obtained following the same steps as of the Einstein's gravity and there is no ambiguity in it \textit{i.e.} the Damour-Navier-Stokes equation can be obtained straightforwardly and there is no alternative picture like the Jordan frame.  Therefore, we start our analysis in the Einstein frame and, thereafter, we move on to discuss the two alternative pictures in the Jordan frame.
\vskip 1mm    
\noindent
{\bf Einstein frame:} 
As we have mentioned earlier, the fluid-gravity correspondence can be obtained from the geometric identity \eqref{NOTDNS1}, which can be written in the Einstein frame as (\textit{i.e.} with the tilde variables)
\begin{align}
\t R_{mn}\t l^m\t q^n_a=\t q^n_a\tliel\t\O_n+\t\th^{(\t l)}\t\O_a-\t D_a(\frac{\t\th^{(\t l)}}{2}+\t\k)+\t D_i\t\sigma^i_a~.
\label{TMNLMQNATIL1}
\end{align}
The LHS of \eqref{TMNLMQNATIL1} can be replaced using the projection $\t E_{mn}\t l^m\t q^n_a$ and, thereby, one can obtain
\begin{align}
8\pi \t T_{mn}^{(\t\phi)}\t l^m\t q^n_a=\t q^n_a\tliel\t\O_n+\t\th^{(\t l)}\t\O_a-\t D_a(\frac{\t\th^{(\t l)}}{2}+\t\k)+\t D_i\t\sigma^i_a~, 
\label{TMNLMQNATIL2}
\end{align}
where, the matter source has been disregarded (\textit{i.e.} the dynamical equation is given as $\t E_{mn}=0$)~. The above equation \eqref{TMNLMQNATIL2} can be identified as the gravitational Navier-Stokes equation or the Damour-Navier-Stokes equation with the following  identifications: $\t F_{a}=\t T_{ab}^{(\t\phi)}\t l^b$ corresponds to the external force term; $\t\pi_a=-\t\O_a/8\pi$ corresponds to the momentum density; the bulk and the shear viscosity coefficients can be identified as $\t\xi=-1/16\pi$ and  $\t\eta=1/16\pi$ respectively (note the total viscous tensor is given as $2\t\eta\t\s^a_b+\t\xi\d^a_b\t\th^{(\t l)}$, where $\t\xi$ corresponds to the bulk viscosity coefficient and $\t\eta$ corresponds to the shear viscosity coefficient); the pressure is identified as $\t P=\t\k/8\pi$~. 

If we had considered the external matter source, the energy-momentum tensor would have appeared into the picture (as the equation of motion, in that case, would be $\t E_{ab}=\t T_{ab}^{(m)}/2$ and, therefore, it would have appeared when one replaces $\t R_{mn}\t l^m\t q^n_a$ of Eq. \eqref{TMNLMQNATIL1} by the projection of $\t E_{mn}\t l^m\t q^n_a$). In that case only the external force term gets modified as $\t F_{a}=(\t T_{ab}^{(\t\phi)}+\t T_{ab}^{(m)})\t l^b$~. Also, note that in the obtained DNS equation \eqref{TMNLMQNATIL2}, there is a Lie derivative of the momentum density. If we change the Lie-derivative by the convective derivative, an extra $\t\th^i_a\t\O_i$ appears on the RHS of\eqref{TMNLMQNATIL2}, which does not have any fluid-dynamic correspondence. This makes DNS equation different from the classical Navier-Stokes equation. This issue has been highlighted in the literature \cite{Price:1986yy, Gourgoulhon:2005ng, Padmanabhan:2010rp}.

Finally, here we define the entropy density as $\t{\mathds S}=\t S/\t A=1/4$ (note the difference between $\t s$ (which has been defined earlier) and $\t{\mathds S}$, though we loosely call both as the ``entropy density''). Thereby, we obtain the shear viscosity coefficient to the entropy density ratio as

 \begin{align}
 \frac{\t\eta}{\t{\mathds S}}=\frac{1}{4\pi}~. \label{KSSEIN}
 \end{align}
 which is the same as of the general relativity and is consistent with the KSS bound (\textit{i.e.} $\frac{\t\eta}{\t{\mathds S}}\geq \frac{1}{4\pi}$).
 
In the following section, we discuss about the fluid-gravity correspondence in the Jordan frame.

\vskip 1mm    
\noindent
{\bf Jordan frame:} Due to the presence of non-minimal coupling in the Jordan frame, the formulation of the fluid-gravity correspondence is a bit non-trivial. As we have mentioned earlier, in this frame, the DNS equation can be obtained in the two different ways and these two ways represents two different representations i) the equivalent picture and ii) the inequivalent picture. In the following, we discuss about the both the pictures and their significance.

\subsection{The inequivalent picture} The DNS equation in the inequivalent picture can be obtained from the identity \eqref{NOTDNS1}, which does not contain any scalar field $\phi$. From a straightforward calculation, one can obtain (incorporating $\phi$)
\begin{align}
\phi R_{mn} l^m q^n_a= q^n_a\liel(\phi\O_n)+\th^{(l)}(\phi\O_a)- \frac{\phi}{2} D_a\th^{(l)}
-D_a(\phi\k)+\phi D_i\sigma^i_a+2l^m q^n_a \Big(\omega_{[m}\na_{n]}\phi\Big)~. \label{ACTDNS1}
\end{align}
The LHS of the above equation \eqref{ACTDNS1} can be replaced by using the projection $E_{mn} l^m q^n_a$ and, thereby, one can obtain
\begin{align}
8\pi \mathcal{T}_{mn} l^m q^n_a= q^n_a\liel(\phi\O_n)+\th^{(l)}(\phi\O_a)- \frac{\phi}{2} D_a\th^{(l)}
-D_a(\phi\k)+\phi D_i\sigma^i_a+2l^m q^n_a \Big(\omega_{[m}\na_{n]}\phi\Big)~, \label{ACTDNSFINAL}
\end{align}
where $\mathcal{T}_{mn}$ is given as
\begin{align}
\mathcal{T}_{mn}=\frac{\o}{8\pi\phi}\Big(\na_m\phi\na_n\phi-\frac{1}{2} g_{mn}\na^i\phi\na_i\phi\Big)-\frac{V g_{mn}}{16\pi}
+\frac{1}{8\pi}\Big(\na_m\na_n\phi- g_{mn}\na_i\na^i\phi\Big)~.
\end{align}
The above equation \eqref{ACTDNSFINAL} has the structure of the DNS-like equation in the Jordan frame. From \eqref{ACTDNSFINAL}, the fluid parameters can be identified as the following: The external force term $F_a$ is identified as $F_a=\mathcal{T}_{ab}l^b$ (for the presence of the external matter field, the force term gets modified as $F_a=(\mathcal{T}_{ab}+T_{ab}^{(m)})l^b$); the momentum density ($\pi_a$) can be obtained as $\pi_a=-\phi\O_a/8\pi$; the pressure term ($P$) is identified as $P=\phi\k/8\pi$; the bulk viscosity coefficient ($\xi$) is obtained as $\xi=-\phi/16\pi$ and the shear viscosity coefficient ($\eta$) can be found as $\eta=\phi/16\pi$. Finally, the last term of Eq. \eqref{ACTDNSFINAL} can be identified as Coriolis-like force term and equation \eqref{ACTDNSFINAL} can be interpreted as the Navier-Stokes equation of a fluid system in a rotating frame, angular velocity of which is provided as $W_a=\na_a\phi/2$. In this case, the ratio of shear viscosity ($\eta$) to the entropy density ($\mathds{S}=\phi/4$) is given as
\begin{align}
\frac{\eta}{\mathds{S}}=\frac{1}{4\pi}~, \label{KSSJOR1}
\end{align}
which saturates the KSS bound and is the same as of the Einstein frame. In addition, this equivalent picture is also consistent with the realization of the non-minimal coupling in terms of the rescaling of the Newtonian constant $G\longrightarrow G_{eff}= G/\phi$ in Einstein's GR \cite{Bhattacharya:2020wdl} \footnote{Here we have considered the geometrized unit and have put $G=1$ throughout. Otherwise, $G$ would have appeared in the expressions of fluid parameters in both the frames. Then it could be shown that the expression of fluid parameters in the Jordan frame, in the inequivalent picture, are identical as of the Einstein's GR except $G$ is replaced by $G_{eff}= G/\phi$. One way of interpreting the non-minimal coupling (in the Jordan frame) is given in terms of the rescaling of the Newtonian constant $G\longrightarrow G_{eff}= G/\phi$ \textit{i.e.} the change of the Einstein-Hilbert Lagrangian $L_{EH}=\mg R/16\pi G \longrightarrow \mg R/16\pi G_{eff}=\mg \phi R/16\pi G$. Our inequivalent picture is consistent with this interpretation.   For details see \cite{Bhattacharya:2020wdl}. }.

As the title implies, the fluid parameters of the Jordan frame, as identified in the inequivalent picture, are not conformally equivalent (as it is explicitly shown by Eq. \eqref{TRANSREL}). Nevertheless, its importance lies in the fact that it is, indeed, a valid way of obtaining fluid-gravity correspondence in the Jordan frame. In addition, it obeys the saturation value of the KSS bound and is consistent with the interpretation of the non-minimal coupling in terms of the rescaling of the Newtonian constant $G$. In the following, it will be shown that an equivalent picture can also be obtained, where the thermodynamic parameters are conformally equivalent. However, in that case, the KSS bound can get violated for some values of $\phi$. In addition, the equivalent picture does not uphold the argument of non-minimal coupling in terms of the rescaling of $G$~.

\subsection{The equivalent picture}
In the previous analysis we have described one way of establishing the fluid-gravity analogy in the Jordan frame and have discussed its theoretical consistency. However the major set back for the inequivalent picture is that the fluid parameters are not conformally equivalent. Therefore, according to the inequivalent picture, there is a frame-dependence in the fluid-gravity correspondence, which is against our general consensus. After all, the entire analysis of obtaining the fluid-gravity analogy is in the classical regime and it is expected that the two frames are classically equivalent (though the two frames can be inequivalent at the quantum level \cite{Kamenshchik:2014waa, Banerjee:2016lco,Ruf:2017xon}). In addition, our analysis for BH thermodynamics in ST gravity also support the equivalence of the two frames. Therefore, the  question arises, whether there is any equivalent picture in the fluid-gravity correspondence of the ST gravity. In the following, we show that an equivalent formalism can also be obtained, where the fluid parameters can be shown to be exact equivalent in the two frames.

The method, which we adopt to obtain the equivalent picture, resembles to that of the section \ref{1STLAWGNC}~. In this case, we obtain the DNS equation in terms of the parameters of the Einstein frame (\textit{i.e.} in terms of $\t\O_a$, $\t\th^{(\t l)}, \t\sigma^b_a$) in the background of the Jordan frame ($g_{ab}$, $l^a$, $q^a_b$, $\nabla_a$, $D_A$, $R_{ab}$, \textit{etc.}).  This requires some involved calculations (for details see \cite{Bhattacharya:2020wdl}), by which one can obtain
\begin{align}
q^n_a\liel\t\O_n-D_a(\frac{\t\th^{(\t l)}}{2}+\t\k)+\t\th^{(\t l)}\t\O_a+D_b\t\sigma^b_a+\t\sigma^i_a(\na_i\ln\phi)=\Big(R_{mn}-\na_m\na_n\ln\phi+\frac{1}{2}(\na_m\ln\phi)(\na_n\ln\phi)\Big)l^mq^n_a~. \label{RELJOR5}
\end{align}
The RHS of \eqref{RELJOR5} can be replaced by the projection $E_{mn} l^m q^n_a$ and, thereby, one can obtain

\begin{align}
q^n_a\liel\t\O_n-D_a(\frac{\t\th^{(\t l)}}{2}+\t\k)+\t\th^{(\t l)}\t\O_a+\frac{1}{\phi} D_b\Sigma^b_a=8\pi \t T_{mn}^{(\t\phi)} l^m q^n_a~,\label{DNSJOr}
\end{align}
where the expression of $\t T_{mn}^{(\t\phi)}$ is given by the Eq. \eqref{TABPHIEQUIV} (which, is equivalent to the expression provided by the Eq. \eqref{TABPHITIL}). The above equation \eqref{DNSJOr} can be identified as the DNS equation of the Jordan frame with the following identifications: the external force term is identified as $F_{a}=\t T_{ab}^{(\t\phi)} l^b$; $\pi_a=-\t\O_a/8\pi$ corresponds to the momentum density; the bulk and the shear viscosity coefficients can be identified as $\xi=-1/16\pi$ and  $\eta=1/16\pi\phi$ respectively and  the pressure is identified as $P=\t\k/8\pi$~. Here $\Sigma^b_a$ corresponds to the shear tensor, which can be identified as $\Sigma^b_a=\phi \t\sigma^b_a$.

With the above identifications, it can be shown that all the fluid parameters are exactly equivalent in the two frames (\ie $\t F_{a}=F_{a}, \t\pi_a=\pi_a$, $P=\t P$, $\xi=\t\xi$, $\t\Sigma_{ab}=\t\sigma_{ab}$ ... {\it etc.}) except for the shear viscosity coefficient, which is connected as $\eta=\t\eta/\phi$. Thus, in this case, the ratio of shear viscosity coefficient to the entropy density is given as
 \begin{equation}
 \label{eta-2}
 \frac{\eta}{\mathds{S}} = \frac{1}{4\pi \phi^2}~,
 \end{equation}
 Thus, the above relation suggests that, in the equivalent picture, the KSS bound ($ \frac{\eta}{\mathds{S}} \ge \frac{1}{4 \pi}$) might be violated for $\phi>1$.
 
 Using AdS-CFT correspondence, it has been argued in literature \cite{Brustein:2008cg} that if any theory, after field redefinition, can be written in terms of Einstein's gravity, the ratio $\eta/\mathds{S}$ will always be equal to $1/4\pi$. Since the ST gravity/ $f(R)$ gravity, after field redefinition, can be written in terms of Einstein's gravity (in the Einstein frame), the argument in the literature \cite{Brustein:2008cg} also implies that the bound should be valid in the ST gravity as well. However, the the aforementioned bound ($ \eta/\mathds{S} = 1/4 \pi$) in the literature \cite{Brustein:2008cg} has been obtained in the specific context of AdS-CFT, whereas our approach is different. Although both the approaches disclose the fluid-gravity correspondence and provides $ \eta/\mathds{S}$ ratio, the concrete connection between the two approaches are not well-defined. Therefore, in future, one needs to explore further along this direction, especially in the context of AdS-CFT, and see which of these two pictures is more accurate. For more discussions, along this line, see \cite{Bhattacharya:2020wdl}.

\section{Conclusions}
As it has been discussed earlier, the thermodynamic interpretation of gravity and the fluid-gravity correspondence provide us with an alternative viewpoint to understand a gravitational theory. In the absence of a proper quantum theory of gravity, these alternative viewpoints turn out to be significantly important. The thermodynamic nature of gravity led the foundation of understanding gravity as ``an emergent phenomena''. Moreover, the fluid-gravity analogy, which was found by Damour led to interpret the $2$-dimensional spacelike section of the event horizon as a fluid bubble. This ``fluid bubble'' viewpoint paved the way for the development of the ``membrane paradigm'' for black holes. Over the years, these topics have been the subject of intense research and have revealed several interesting facts on the nature of gravity. Thus, the analysis of both the thermodynamic and the fluid-gravity analogies have the paramount importance to understand a gravitational theory.

 In this review, we discussed these two alternative viewpoints (\textit{i.e.} thermodynamic interpretation and fluid-gravity correspondence) for the scalar-tensor gravity, which is considered as one of the most potential candidates among several extensions of GR. The study is highly non-trivial for two main reasons, (i) the presence of the non-minimal coupling in the action of the Jordan frame and (ii) the issue of physical equivalence/ in-equivalence of the Jordan and Einstein frames, which has been debated over the years. Here we have provided a complete picture of both thermodynamic and fluid-gravity correspondence analogies, which has been obtained in fragments in our previous works. 
 
 Here, the scalar-tensor theory of gravity has been studied extensively starting from the action level. It has been shown that the action in the two frames are equivalent only up to a total derivative term (\textit{i.e.} the $\square\phi$ term). Later, it has been found that the the holographic nature of the Einstein-Hilbert action is missing in the Jordan frame, albeit present in the Einstein frame. This makes the two frames in-equivalent at the action level itself, which has been found later to culminate in the in-equivalence of the two frames at the thermodynamic level (which is, of course, removed by incorporating several assumptions into the account \cite{Koga:1998un}). To obtain the holographic relation in the Jordan frame, it has been found that the $\square\phi$ term is required to be incorporated in the action of the Jordan frame, which has been left out in literature. In addition, it has been found that the condition, which restore the holographic nature in the Jordan frame (\textit{i.e.} the inclusion of the $\square\phi$ term) also establishes the thermodynamic equivalence in the two frames. Thus, the holographic relation plays a pivotal role in establishing the exact thermodynamic equivalence in the two frames, which went unnoticed in the earlier works.
 
 Although the entire thermodynamic description approve the equivalence of the two frames, the fluid-gravity analogy presents both the equivalent as well as the in-equivalent picture. Therefore, further investigation is required along this line to obtain a concrete understanding. We hope to contribute soon in this direction.
 \vskip 3mm
 \section*{Acknowledgement}
 This work is inspired by the works of Prof. T. Padmanabhan and, we dedicate this paper to his memory. The work of one of the authors (B.R.Majhi) is supported by Science and Engineering Research Board (SERB), Department of Science \& Technology (DST), Government of India, under the scheme Core Research Grant (File no. CRG/2020/000616). 
 \vskip 3mm
 \textbf{Data availability:} This manuscript has no associated data or the data will not be deposited.

\end{document}